\pdfoutput=1

\documentclass[fleqn, useAMS, usenatbib]{mnras}
\usepackage{times}
\usepackage{graphicx}
\usepackage{dcolumn}
\usepackage{amsmath}
\usepackage{bm}
\usepackage{amssymb}
\usepackage{natbib}
\usepackage{subfig}
\usepackage{subfig}

\hypersetup{pdfauthor={D. A. Green}, pdftitle={The Chaotic Long-term X-ray Variability of 4U 1705--44},
               pdfkeywords={stars: neutron -- stars: individual: 4U1705--44 -- X-rays: binaries -- chaos -- accretion, accretion discs -- methods: numerical},
               bookmarksnumbered=true}

\makeatletter
 \def\@textbottom{\vskip \z@ \@plus 1pt}
 \let\@texttop\relax
\makeatother

\title[4U1705--44 Long-term Variability]{The Chaotic Long-term X-ray Variability of 4U 1705--44}

\author[R. A. Phillipson et al.]
  {R.\,A.~Phillipson,$^1$\thanks{E-mail: rebecca.a.phillipson@drexel.edu}
  P.\,T.~Boyd,$^2$\thanks{E-mail: patricia.t.boyd@nasa.gov} A.\,P.~Smale,$^2$\\
  $^1$Department of Physics, Drexel University, 3141 Chestnut St, Philadelphia, PA 19104, USA\\
  $^2$Astrophysics Science Division, NASA Goddard Space Flight Center, Greenbelt, MD 20771, USA}
\date{Accepted 2018 April 15. Received 2018 March 26; in original form 2017 August 10\\
Accepted version of R A Phillipson et al. The chaotic long-term X-ray variability of 4U 1705-44. MNRAS (2018) 477 (4): 5220-5237. By permission of Oxford University Press on behalf of the Royal Astronomical Society. Published online at: https://academic.oup.com/mnras/article/477/4/5220/4975798 For permissions, please email journals.permissions@oup.com}

\pubyear{2017}
\pagerange{\pageref{firstpage}--\pageref{lastpage}} \pubyear{2002}

\def\LaTeX{L\kern-.36em\raise.3ex\hbox{a}\kern-.15em
    T\kern-.1667em\lower.7ex\hbox{E}\kern-.125emX}

   \volume{{\rm in press}}

\begin{document}

\label{firstpage}

\maketitle
\begin{abstract}
The low-mass X-ray binary 4U1705--44 exhibits dramatic long-term X-ray time variability with a timescale of several hundred days. The All-Sky Monitor (\textit{ASM}) aboard the \textit{Rossi X-ray Timing Explorer} (\textit{RXTE}) and the Japanese \textit{Monitor of All-sky X-ray Image} (\textit{MAXI}) aboard the International Space Station together have continuously observed the source from December 1995 through May 2014. The combined \textit{ASM-MAXI} data provide a continuous time series over fifty times the length of the timescale of interest. Topological analysis can help us identify 'fingerprints' in the phase-space of a system unique to its equations of motion. The Birman-Williams theorem postulates that if such fingerprints are the same between two systems, then their equations of motion must be closely related. The phase-space embedding of the source light curve shows a strong resemblance to the double-welled nonlinear Duffing oscillator. We explore a range of parameters for which the Duffing oscillator closely mirrors the time evolution of 4U1705--44. We extract low period, unstable periodic orbits from the 4U1705--44 and Duffing time series and compare their topological information. The Duffing and 4U1705--44 topological properties are identical, providing strong evidence that they share the same underlying template. This suggests that we can look to the Duffing equation to help guide the development of a physical model to describe the long-term X-ray variability of this and other similarly behaved X-ray binary systems. 
\end{abstract}


\section{Introduction} \label{sec:intro}

X-ray binaries exhibit periodicities on multiple timescales, which provide information about the physical mechanisms at play. 
High magnetic fields and disc-magnetosphere interactions create pulsations on timescales from milliseconds to seconds \citep{vanDerKlis}. 
Orbital modulations are seen from minutes to tens of days \citep{Levine2011}. 
A number of X-ray binaries show evidence of long-term periodicities, or super-orbital periods, on timescales much longer than their orbital periods (\citealt{Clarkson1}, \citealt{Clarkson2}). 
Some X-ray binaries have variability on timescales of over a hundred days that are not strictly periodic \citep{Boyd2000}. 

For low-mass X-ray binaries (LMXBs), it is generally accepted that the accretion is primarily due to a substantial accretion disc, which is unstable to irradiation-driven warping \citep{Foulkes2}. The precession of the accretion disc is potentially the mechanism underlying the observed long periods. Warped or twisted accretion discs in general are also invoked to explain phenomena observed across many types of systems including high-mass X-ray binaries, cataclysmic variables, proto-planetary discs and active galactic nuclei \citep{Ogilvie2}. One relevant example is SS433, which has a measured 160-day precession of the relativistic jets identified to be associated with precession of the accretion disc about the central neutron star (\citealt{Whitmire}; \citealt{Collins}). Other well-known examples in X-ray binary systems include Her X-1, with a varying long-term period of about 35 days associated with its observed anomalous low states \citep{Still}, and SMC X-1, with confirmed 60-day variations in the X-ray \citep{Wojdowski}.

In this study, we consider the low-mass X-ray binary, 4U1705--44, which exhibits the high-amplitude transitions and non-periodic long-term variability of interest ( \citealt{Hasinger}; \citealt{Muno}). 4U1705--44 is a neutron star with mass $1.1 - 1.6 \, M_\odot$ \citep{Olive} at a distance of $7.4 \,kpc$ \citep{D'Ai}. It is considered to be part of the class of 'atoll' sources \citep{Wang}, however it has sampled the upper parts of a Z in the X-ray colour-colour diagram during its soft-to-hard state transitions \citep{Barret}. An infrared counterpart has been observed suggesting a dwarf star companion with a 1-10hr orbital period \citep{Homan}, corresponding to a mass less than 0.5 solar masses. 

The \textit{RXTE} (\textit{Rossi X-ray Timing Explorer}) All Sky Monitor obtained approximately 14 years of daily monitoring in the 2-12 keV energy range of 4U1705--44 with its scan of 80 per cent of the sky every 90 minutes \citep{Bradt}. \textit{MAXI} (\textit{Monitor of All-Sky X-ray Image}) on-board the ISS, in operation since August 2009, scans the sky every 96 minutes in the 2-20 keV band and continues to provide daily monitoring on 4U1705--44 \citep{Matsuoka}. With \textit{MAXI} and \textit{RXTE} combined, we have a nearly 20 year continuous light curve revealing over 50 cycles of the high amplitude transitions that occur on timescales on the order of hundreds of days. It is only recently that such continuous, high signal-to-noise, uninterrupted time series have been available at the timescales of interest for investigating long term variability in X-ray binaries on the order of years. The long-term modulations, which have also been observed in some other LMXBs, are not strictly periodic \citep{Boyd2004}. In this paper, we present evidence that the long-term variability is reminiscent of a nonlinear double-welled oscillator evolving in the chaotic regime. We find that the Duffing oscillator is a good candidate to describe the system. Since the parameter ranges that best represent the 4U1705--44 light curve are indeed in the chaotic regime of the Duffing oscillator, we use analyses appropriate for non-linear and chaotic time series.

There are two broad approaches to understanding chaotic behaviour in dynamical systems. The metric approach generally involves computing the Lyapunov exponents and various dimensions, such as the Correlation and Minkowski dimensions, giving a tight range for the fractal dimension (\citealt{Ott}; \citealt{Anishchenko}). However, these methods require very large datasets and degrade rapidly with noise and are therefore rarely applicable to astronomical data \citep{Mindlin1}. The topological approach involves the identification of the two mechanisms that are responsible for the creation of a strange attractor. These two mechanisms are the stretching and squeezing mechanisms correlating to the sensitive dependence on initial conditions and recurrence phenomenon, respectively \citep{Mindlin2}. The stretching and squeezing mechanisms can be characterised by computing how the unstable periodic orbits are uniquely organized topologically \citep{Letellier}. In fact, it is possible to determine how the unstable periodic orbits embedded in the attractor are organized in terms of a set of integers (\citealt{Solari1}, \citealt{Tufillaro}). Extracting these integers from a time series is robust against noise and can be performed for smaller datasets \citep{GilmoreTopologyBook}.

Thus, we will follow the topological approach. Two of the topological invariants are called the linking numbers, the unique set of integers aforementioned, and the relative rotation rates (RRR), a set of fractions which when summed over all initial conditions equal the linking numbers \citep{Gilmore}. It is straightforward and sufficient to compute the RRR \citep{Gilmore}, which we will do for both 4U1705--44 and the Duffing oscillator \citep{Solari2}. By extension of the Birman-Williams theorem (\citealt{Birman2}; \citealt{Birman1}), two chaotic attractors are equivalent if they are described by branched manifolds that can be smoothly deformed, one into the other. In other words, the RRR and linking numbers will be invariant under transformations and control-parameter changes. Conversely, the 4U1705--44 and Duffing oscillator time series share the same underlying template if their RRR are identical. Identifying the underlying template of 4U1705--44 via the RRR allows for the prediction of the RRR of all other possible orbits in the time series and provides a qualitative model for the flow that uniquely generates the chaotic times series (\citealt{Mindlin1}; \citealt{Letellier}). If both the 4U1705--44 and Duffing oscillator time series share the same underlying template, then we can look to the Duffing oscillator equation and its template to make qualitative predictions of the behaviour of 4U1705--44 and infer possible physical parameters of the system \citep{GilmoreTopologyBook}.

To follow the topological analysis procedure we first must determine the lowest order period in the data in question. We can then use the close returns method as used in \cite{Boyd94} to locate and extract regions in the time series that can be used as surrogates for the unstable periodic orbits. The topological invariants of RRR of all pairs of unstable periodic orbits extracted can then be computed from the time series and compiled into a matrix or table, as exemplified by \cite{Solari2} for the Duffing oscillator. This procedure was laid out generically for the computation of RRR by \cite{Mindlin2} and \cite{Gilmore}.

This paper is organized as follows. In Section \ref{sec:Data} we will review the data obtained from 4U1705--44 using the \textit{RXTE ASM} and \textit{MAXI} and our data reduction techniques, as well as the generation of numerical times series from the Duffing oscillator equations of motion. In Section \ref{sec:Period} we will determine the lowest order period of an unstable periodic orbit for both the 4U1705--44 and Duffing data using the close returns method and via dynamical power spectra and time-delay embedding analyses. In Section \ref{sec:Extract} we will use the lowest order period and close returns method to locate and extract regions in the time series that can be used as surrogates for the unstable periodic orbits. In Section \ref{sec:RRR} we will compute the topological invariants of RRR of all pairs of periodic orbits extracted from the time series and compile these RRR into matrices for the 4U1705--44 and Duffing data. Finally, in Section \ref{sec:Conclusions}, we will compare the resulting intertwining matrices containing the RRR followed by a discussion and concluding remarks.

\section{Data and Methods} \label{sec:Data}

\subsection{4U 1705--44}
The \textit{Rossi X-ray Timing Explorer} (\textit{RXTE}) satellite was launched December 30, 1995 and was decommissioned January 5, 2012 \citep{Bradt}. The \textit{RXTE} satellite embodied three instruments to make observations in the X-ray bandwidth (2-200 keV) with a large effective collecting area (~$0.8\, m^2$). One of the instruments was the All-Sky-Monitor (\textit{ASM}), which scanned 80 per cent of the sky every 90 minutes in the 2-12 keV energy band, with a timing resolution of 90 minutes or more and sensitivity of 30 mCrab \citep{Levine}. The low-mass X-ray binary system 4U1705--44 was one of the ~75 sources continuously observed by \textit{RXTE}'s \textit{ASM}, and was immediately noted by \cite{Levine} to have high-amplitude variability ranging from $<$25 mCrab up to 300 mCrab. 

The \textit{Monitor of All-sky X-ray Image} (\textit{MAXI}) telescope is aboard the International Space Station (ISS) orbiting Earth as part of the Japanese Experiment Module with the goal of obtaining a map of the X-ray sky with greater sensitivity than previous all-sky monitors. \textit{MAXI} has been in operation since August 2009 and scans nearly the entire sky every 96 minutes in the 2-20 keV energy band and with a sensitivity of 20 mCrab \citep{Matsuoka}. \textit{MAXI} obtained observations of 4U1705--44 since its operation start. We therefore have continuous, daily observations of 4U1705--44 and other bright X-ray sources from December 1995 through the present, sampling more than two decades. For this investigation, we consider the light curve of 4U1705--44 from 6 January 1996 through 16 May 2014. Furthermore, because \textit{MAXI} and \textit{RXTE ASM} were in operation simultaneously for 3 years, the light curves of 4U1705--44 can be combined in a seamless manner.

\begin{figure*}
	\centering
	\includegraphics[width=0.9\textwidth]{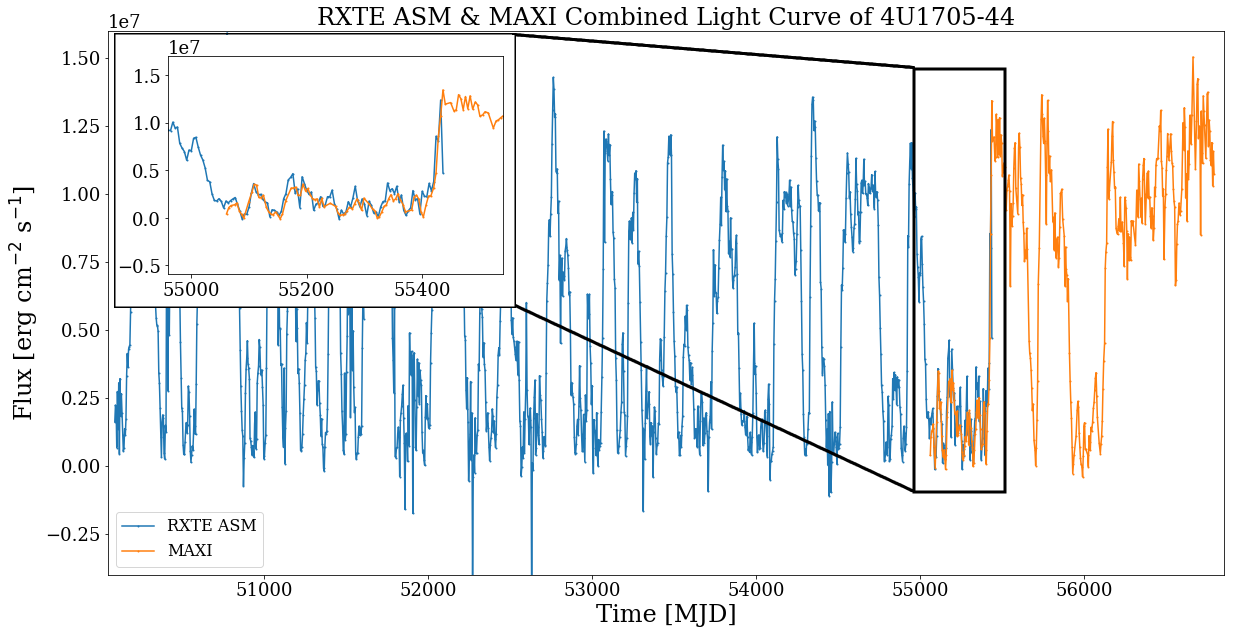}	
	\caption{Flux of 4U1705--44, 6 January 1996 to 16 May 2014, in MJD. The blue curve represents the data obtained from \textit{RXTE ASM}, while the orange curve is from \textit{MAXI} with data taken in the 2-12 keV and 2-20 keV flux bands, respectively. The region in which the blue and orange curves overlap (inset) provides the basis for scaling the \textit{MAXI} time series to that of \textit{RXTE}.}
	\label{fig:rawdata}					
\end{figure*}

\begin{figure*}
	\centering
	\includegraphics[width=0.9\textwidth]{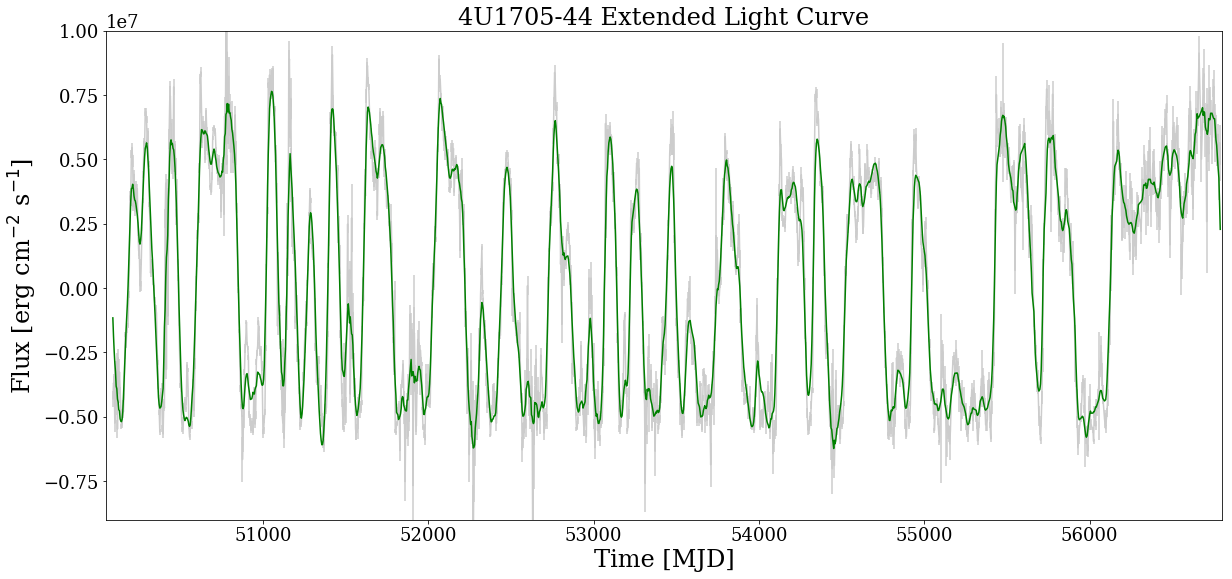}	
	\caption{The mean-subtracted flux of 4U1705--44, 6 January 1996 to 16 May 2014 (in MJD), using the combined \textit{RXTE ASM} and \textit{MAXI} data in physical flux units via normalisation to \textit{Crab} units of each instrument. The grey curve is the raw light curve and associated error. The green curve was produced by applying a low-pass filter.}
	\label{fig:timeseries}					
\end{figure*}

\begin{figure}
	\centering
	\includegraphics[width=0.45\textwidth]{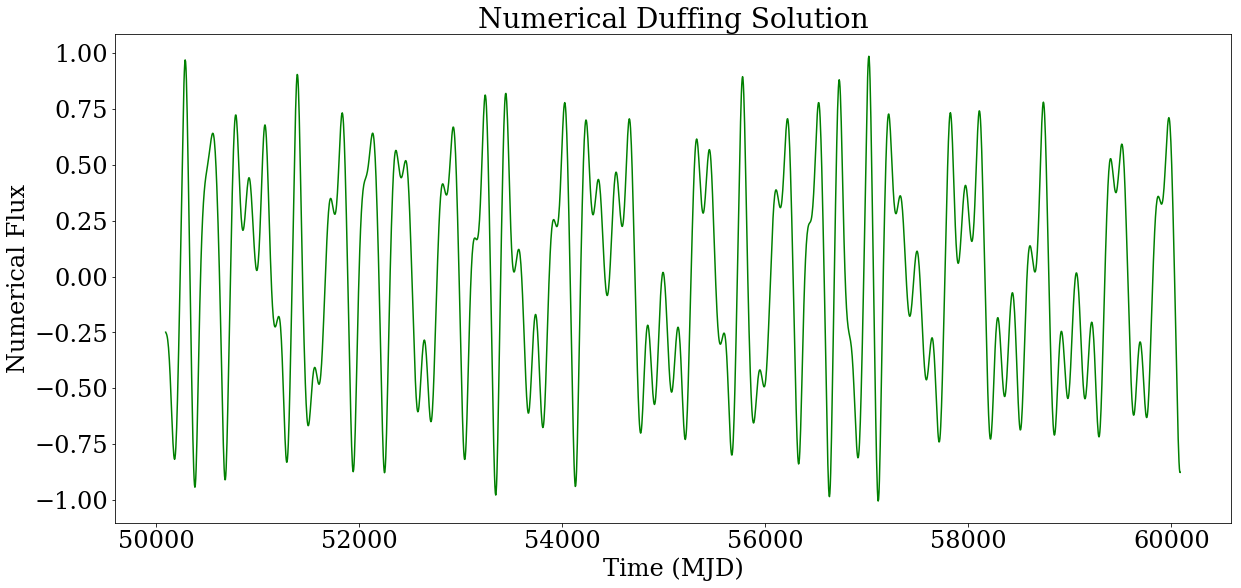}	
	\caption{Duffing Equation solution time series segment spanning the same time window as the 4U1705--44 extended light curve.}
	\label{fig:dufftimeseries}					
\end{figure}

The construction of the 4U1705--44 time series consisted of combining \textit{RXTE ASM} and \textit{MAXI} data. For both datasets, we re-binned the data to every 4 days, since our primary timescale of interest was over hundreds of days, thereby improving our signal-to-noise. The online tool\footnote{www.dsf.unica.it/$\sim$riggio/calcs.html} based on \cite{Kirsch} converts from a flux expressed in Crab flux units in a specific energy band to physical units of erg cm$^{-2}$ s$^{-1}$ for ease of reporting the flux from various instruments in terms of the Crab nebula. Given the respective average count rate from the Crab nebula observed by each instrument over the same observation time as 4U1705--44 ($75.3$ counts/second for \textit{RXTE ASM} and $3.25$ counts/second for \textit{MAXI}) and the corresponding physical units from both instruments in each bandwidth (1.5 - 12 keV and 2 - 20 keV) corresponding to 1 Crab, we can therefore report the light curve in terms of the physical units taking into consideration the differing energy bands of \textit{RXTE ASM} and \textit{MAXI}. The same method of normalising to Crab units to form an extended light curve in physical units was used for the multi-wavelength, multi-instrument analysis of the X-ray binary, LMC X--3, by \cite{Torpin}. 

Our methods will also require an evenly sampled dataset. Following the same procedure as \cite{Smale}, any gaps or values that contained errors greater than $5\sigma$ were replaced by a random value containing the same mean and standard deviation as the entire dataset. This treatment was applied to 4.8 per cent of the data.

The final year of observations from the \textit{RXTE ASM} had consistently higher noise, which was seen in many light curves from various sources as also noted by \cite{Smale}. In order to avoid potential contamination of the time series, we determined a cut-off point for the \textit{ASM} data. As both the Crab Nebula and Cassiopeia A are bright sources with a consistently high signal-to-noise, we chose the cut-off point to be where the errors in the daily measurements of these two sources consistently exceeded the $3\sigma$ range in the same daily monitoring of 4U1705--44, which occurred in October of 2010, in agreement with the cut-off date determined by \cite{Smale}.

\textit{MAXI} started its observations of 4U1705--44 in 2009, before the October 2010 ASM cut-off date, and we could therefore use the overlap to appropriately scale the \textit{MAXI} data to the \textit{ASM} data. That is, starting with the two light curves displayed in Fig.~\ref{fig:rawdata}, we used the light curve from \textit{RXTE ASM} as a reference and scaled the entire light curve from \textit{MAXI} such that the amplitude range (maximum and minimum flux difference) in the region in which both light curves overlapped were equal. Finally, we normalised the entire combined dataset by the maximum amplitude of flux present in the light curve before applying all methods described below.

The resulting time series, showing both the raw data with errors and a low-pass filtered version (in order to smooth the high-frequency oscillations) is plotted in Fig.~\ref{fig:timeseries}. It is clear by visible inspection of Fig.~\ref{fig:rawdata} and Fig.~\ref{fig:timeseries} that, while the light curve displays large amplitude oscillations where the flux increases by more than a factor of ten from minimum to maximum, these oscillations are far from strictly periodic. In addition to the large amplitude oscillations, the flux from 4U1705--44 displays smaller amplitude oscillations around the low flux level and the high flux level. In general, the longer the system stays near the average high(low) flux level, the more smaller amplitude oscillations are seen. This is reminiscent of the one-dimensional time series from the classic Duffing equation, a canonical nonlinear oscillator shown in Fig.~\ref{fig:dufftimeseries}. This similarity motivates us to more quantitatively compare the 4U 1705-44 light curve to the Duffing oscillator.

\subsection{Chaos and Topology}

\subsubsection{Is the Time Series of 4U 1705--44 Chaotic?}

\begin{figure*}
	\centering
	\includegraphics[width=0.9\textwidth]{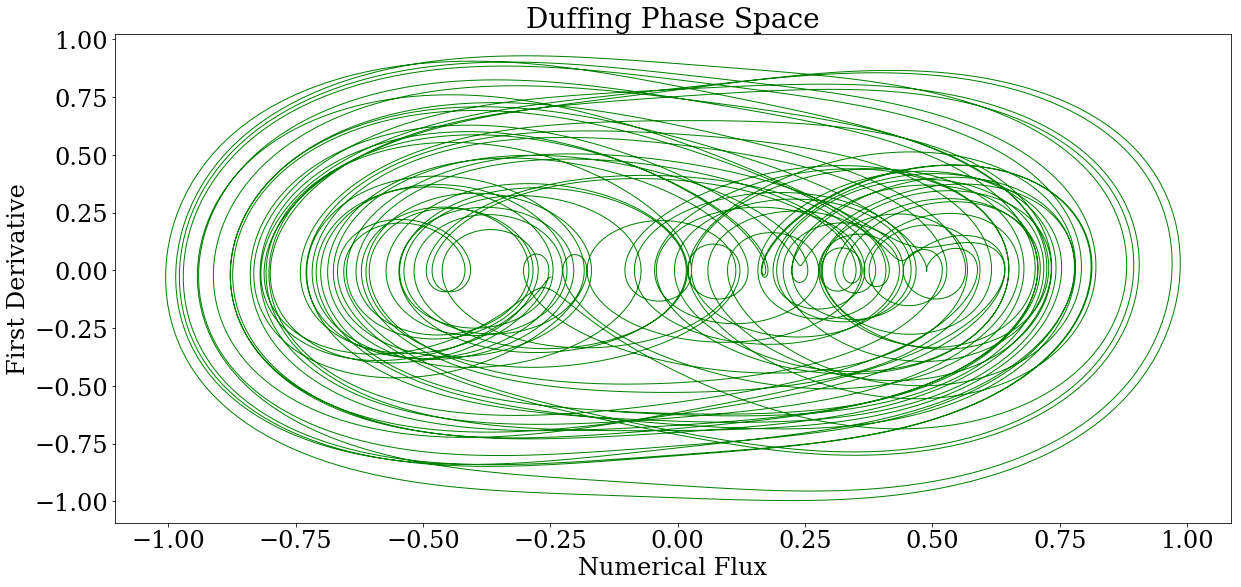}	
	\caption{Numerical Flux of Duffing Equation solution against its first derivative, normalised to unity. We note the similar double-welled behaviour in phase space similar to 4U1705--44, and the slight asymmetry in tightness in the two wells.}
	\label{fig:duffphasespace}					
\end{figure*}

The low-mass X-ray binary, 4U 1705--44, has previously been found to have high-amplitude transitions not easily fitted by a simply periodic model but with a power spectrum that displays clusters of multiple periodicities \citep{Durant}, characteristics often exhibited by both chaotic and stochastic nonlinear systems. Signatures of deterministic chaos have also been detected as the driver of observed variability in five black hole X-ray binaries \citep{Sukova} on short timescales relating to quasi-periodic flares and oscillations. Unlike stochastic systems, low-dimensional chaotic systems display characteristic behaviour when cast into a phase space (for example, by plotting the flux-like variable and its first time derivative; \cite{Takens}; \cite{Ott}). The phase space embedding of the Duffing oscillator time series (Fig.~\ref{fig:dufftimeseries}) is shown in Fig.~\ref{fig:duffphasespace}. The two centres in phase space about which the trajectory evolves are the hallmark of this 'double-well potential' \citep{Holmes}. We discuss the Duffing oscillator in more detail in Sec.~\ref{sec:duff}.

\begin{figure*}
	\centering
	\includegraphics[width=0.9\textwidth]{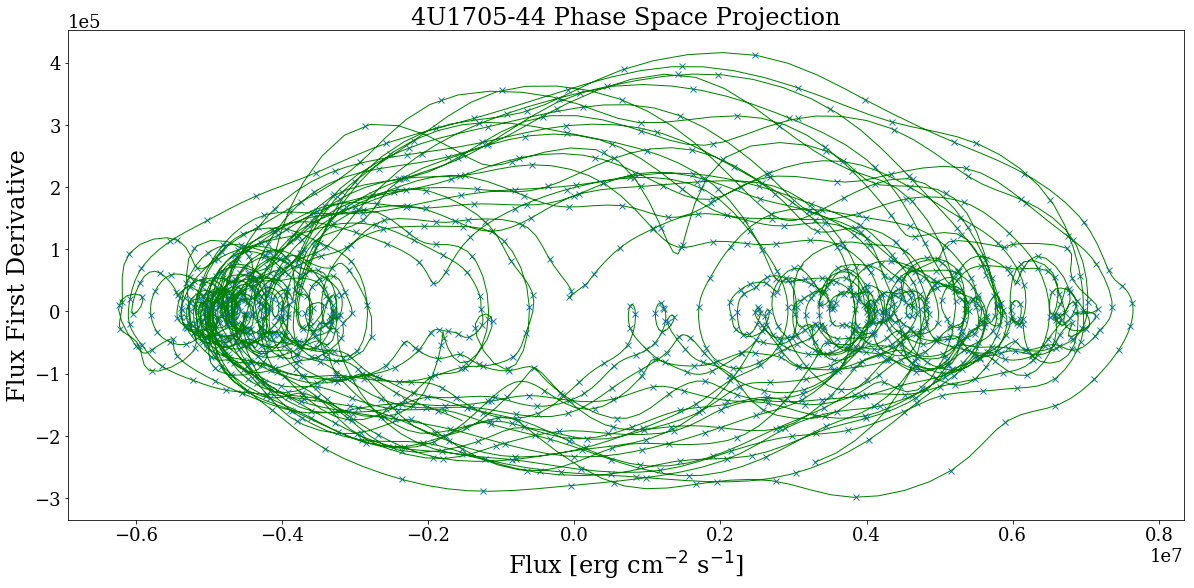}	
	\caption{The mean-subtracted flux of 4U1705--44 against its first derivative, over the time period of December 1995 to May 2014 using combined \textit{RXTE ASM} (1.5-12 keV) and \textit{MAXI} (2-20 keV) data. The green curve represents a cubic-spline interpolation of the data (blue crosses).}
	\label{fig:phasespace}					
\end{figure*}

The phase space trajectory (the flux versus its first time derivative) of 4U1705--44 is shown in Fig.~\ref{fig:phasespace}. Upon initial visual inspection of the phase-space embedding (Fig.~\ref{fig:phasespace}), we see the first indication of a chaotic structure, i.e. an axis about which the flow evolves, or a 'hole' in the middle of the projection \citep{Letellier}, which indicates the presence of (unstable) periodic orbits. The 4U1705--44 phase space embedding also shows evidence of two centres of rotation suggestive of a double-well potential.

Distinguishing between chaotic or stochastic driven time series is an ongoing endeavour in time series analysis in general and particularly challenging in the astrophysical context due to the often short and unevenly sampled data. Chaotic time series arising from highly nonlinear systems produce similar trends to time series arising from stochastic processes, namely, the long-term unpredictable behaviour and broad-band power spectra \citep{Osborne}. Second-order metrics, such as the power spectrum or autocorrelation function, do not contain enough information to distinguish between the underlying processes; a nonlinear stochastic process and a deterministic chaotic process can have the same statistics \citep{Zunino}. Furthermore, non-stationarity in a time series can also result in falsely identifying nonlinear structures \citep{Schreiber2000}. A new trend of utilising higher-order metrics, such as time reversibility and third-order autocorrelation functions, has therefore developed in order to explore the presence of nonlinearity and determinism (\citealt{Collis}; \citealt{Gao}; \citealt{Zunino}).

For data driven analysis, the application of nonlinear time series methods should be justified by establishing nonlinearity in the time series. Using the method of surrogate data \citep{Theiler1992}, we can determine to a significance level of 95 per cent that nonlinearity is present in the 4U1705--44 time series by testing for time reversal asymmetry in the 4U1705--44 light curve (\citealt{Diks}; \citealt{Schreiber2000}). Furthermore, the local vs. global linear prediction test by \cite{Casdagli1992} and recurrence analysis (\citealt{Casdagli1997}; \cite{Marwan}) also indicate that the 4U1705--44 time series is likely deterministic and chaotic. See the Appendix for a full description of the surrogate data method, nonlinearity tests, and recurrence analysis. We will therefore use nonlinear analysis and chaos topology methods to characterise the variability of the 4U1705--44 light curve.

\subsubsection{Topology Methods}

At the center of chaos theory are two fundamental properties of an attractor. First is the 'stretching' mechanism, or the sensitive dependence on initial conditions. This can be mathematically characterised by the exponential divergence between nearby trajectories in the phase space of an attractor, where the exponent of the diverging term is called the Lyapunov exponent and must be positive. The 'squeezing' mechanism relates to the characteristic that the trajectories in the phase space are bounded, corresponding to negative Lyapunov exponents. The constant stretching and squeezing of trajectories constitutes the fractal behaviour of a chaotic attractor (\citealt{Gao}, \citealt{Ott}).

There are various ways to visualise an attractor beyond determining the Lyapunov exponents or fractal dimension. Poincar{\'e} developed topology (the compilation of his original papers from 1892 through 1904, translated into English, are available in \cite{PoincareBook}) in order to study differential equations geometrically. This resulted in Poincar{\'e} sections, bifurcations, and, with the entrance of Lorenz, more complex return maps, all of which are related to the phase space of an attractor (\citealt{Lorenz}; \citealt{Ott}; \citealt{Anishchenko}; \citealt{PoincareBook}). In the past thirty years, topological methods from Lie group theory have also been used to study low-dimensional chaotic attractors (\citealt{Gilmore}; \citealt{GilmoreTopologyBook}; \citealt{Letellier}). Poincar{\'e} identified chaotic attractors as recurrent but not periodic (the Poincar{\'e} recurrence theorem). That is, a chaotic trajectory will return to the neighbourhood of an initial position given enough time. Trajectories that are then perturbed just enough to close on themselves are periodic orbits, but are not stable. The mechanism(s) that creates the attractor simultaneously creates and organizes all the unstable periodic orbits in it. The advantage of the topological methods drawn from group theory is in their ability to then apply an integer invariant associated with each pair of closed, unstable orbits, called a linking number. Topological analysis using templates and bounding tori (\citealt{Birman2}; \citealt{Birman1}; \citealt{Mindlin1}; \citealt{Mindlin3}; \citealt{Mindlin2}) extends the geometrical methods of Poincar{\'e} to characterise and classify chaotic attractors embedded in three-dimensional spaces \citep{Letellier}. The first branched manifold for describing a chaotic attractor was drawn in 1963 by Lorenz using 'isopleths' \citep{Lorenz}.

The connection between the embedded phase space of a chaotic attractor and a template (branched manifold, or 'knot-holder') depends on the Birman-Williams theorem, which states that topological invariants of the periodic orbits are the same in the attractor as in its so-called 'caricature', the two-dimensional branched manifold (\citealt{Birman2}; \citealt{Birman1}; \citealt{Gilmore}). The transformation from a chaotic attractor embedded in a three-dimensional phase space to a branched manifold or template and back preserves the topological invariants (e.g. linking numbers, relative rotation rates). Furthermore, these invariants are unique to specific branched manifolds. For example, the canonical Lorenz attractor corresponds to a variation of the Smale horseshoe template \citep{Rossler1977} whilst the R{\"o}ssler system, considered the simplest chaotic attractor from the topological point of view, corresponds to a simple stretched and folded ribbon \citep{Rossler1976}.

\textit{The crucial point relevant to our analysis is the fact that these two chaotic attractors are topologically inequivalent because their underlying templates are different and thus their unstable periodic orbits are organized differently, as described by the difference in their linking numbers and relative rotation rates}. In fact, the three most widely cited examples of low-dimensional dynamical systems exhibiting chaotic behaviour, the Lorenz, R{\"o}ssler, and van der Pol-Shaw attractors, are all associated with different branched manifolds, and are therefore intrinsically inequivalent (\citealt{Gilmore}; \citealt{GilmoreBook}). 

The Birman-Williams theorem becomes especially powerful in the comparison of topological invariants for periodic orbits extracted from data with the invariants of corresponding orbits in a branched manifold. The equivalence of two underlying topological templates does not necessarily dictate that they look identical when projected onto two dimensions, such as in a time series or embedded phase space. Rather, by identifying the underlying template through the determination of its topological invariants, one can make a mathematically rigorous statement about whether two systems are related, or not (\citealt{Solari1}; \citealt{Mindlin1}; \citealt{Tufillaro}).

In this paper, we seek to extract and compare the topological invariants of two different time series, one from the light curve of 4U1705--44 and the other from the numerical solution of the Duffing oscillator. By the Birman-Williams theorem, the equivalence of the invariants (or lack thereof) will directly determine the relationship between the two underlying templates that organize the unstable periodic orbits on the attractor. If we find that the templates are identical (or, more accurately, not inequivalent) then we can make the case that the equations of motion producing the two different time series are related.

\subsection{The Duffing Oscillator}\label{sec:duff}

The phase space trajectory of 4U1705--44 displays a double-welled behaviour where one well is more favoured and tightly wound than the other. Such behaviour is shared by some damped and driven harmonic oscillators. We therefore chose to explore various nonlinear oscillators that could display double-welled behaviour similar to 4U1705--44 and compare their characteristics. This led us to study the Duffing oscillator, which within certain parameter regimes remarkably resembles the light curve and phase space trajectory of 4U1705--44. With the goal of performing a parallel topological analysis on the 4U1705--44 data and generated solutions of the Duffing equation in mind, we systematically solved the Duffing equation with randomly varying parameter ranges in order to produce many solutions for comparison.

A generic form involving five parameters of the Duffing equation was used:
\begin{equation}\label{eq:duff}
	\ddot{x} = \delta \dot{x} + \beta x^3 - \alpha x + \gamma \cos{\omega t}.
\end{equation}

Solutions were generated using a 4th-order Runge-Kutta method within randomly sampled parameter ranges for which chaotic behaviour arises. Those solutions with qualitatively similar frequency of low-order, low-high amplitude transitions and double-welled phase space features whereby one well was favoured over the other predominated an increasingly narrow range in parameter space. The size of the damping is controlled by $\delta$, the non-linearity of the restoring force by $\beta$, the stiffness of the oscillator by $\alpha$, the amplitude of the forcing by $\gamma$ and the driving frequency by $\omega$. Of particular interest, the $\omega$ driving parameter frequented the range between $3.91$ and $5.15$, corresponding to driving periods of 122 to 160 days. The Duffing solution plotted in phase space in Fig.~\ref{fig:duffphasespace} corresponds to the parameter values ($\beta =  4.99$, $\alpha = 8.18$, $\delta = 0.31$, $\gamma = -6.81$, and $\omega = 4.01$), with a driving period of 140 days. These representative parameter values allow for a double-welled oscillator with moderate chaotic behaviour and is the solution we used as the reference Duffing solution for our topological analysis in this paper. (For a comparison, the parameter values often used to study the chaos of the Duffing oscillator are ($\beta =  1$, $\alpha = 1$, $\delta = 0.2$, $\gamma = -0.3$, and $\omega = 1$).) We chose the same sampling rate as 4U1705--44 and normalised the resulting numerical solutions for comparison to the combined \textit{ASM-MAXI} data of 4U1705--44. Our numerical solutions are ten times the length of the 4U1705--44 time series.

Note that our goal is not to generate a numerical time series from the Duffing oscillator that exactly reproduces the X-ray binary light curve; since sensitive dependence on initial conditions is a hallmark of chaos, this would not be possible. Rather, our goal is to investigate whether the two time series share the same topological invariants, as would be expected if they both are low-dimensional chaotic systems evolving according to the same family of equations of motion. To do this comparison requires extracting unstable periodic orbits from both the real and the numerical time series, in order to identify the underlying templates, and compare them to each other topologically. For this, we chose a segment of the simulated time series that qualitatively reproduces the same behaviour in phase space, sufficient to perform this analysis.

\section{Finding the Lowest Order Period of 4U1705--44}\label{sec:Period}
\begin{figure*}	
	\centering
	\includegraphics[width=0.9\textwidth]{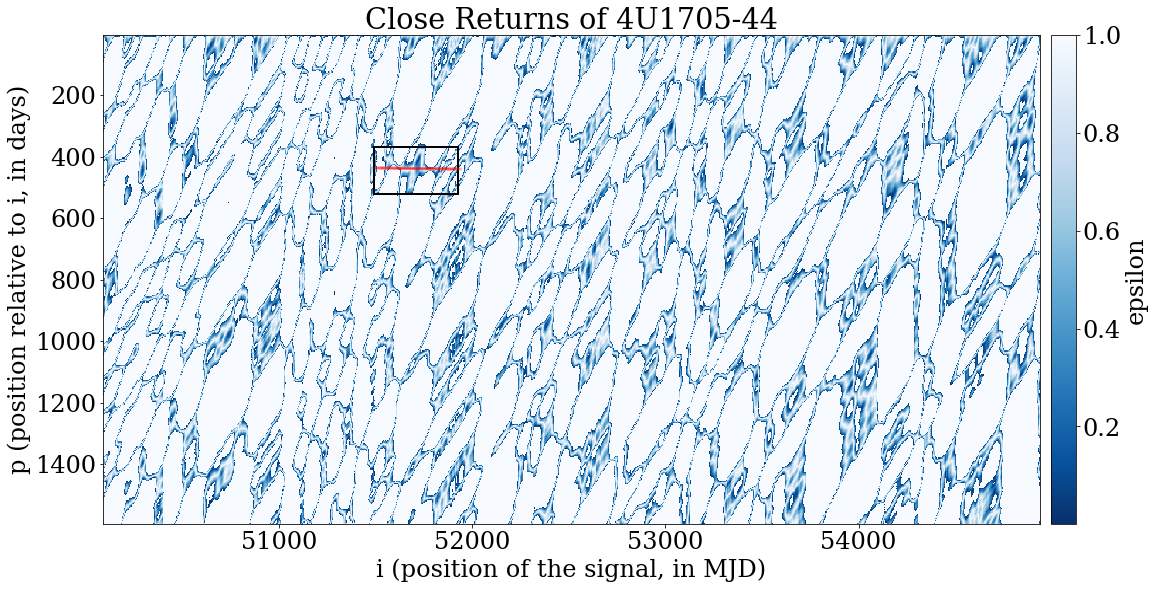}	
	\caption{Close Returns of 4U1705--44: A blue point is plotted at $(i, p)$ when, for that position in time, $i$, there is a point $p$ days later that is 'close' to the flux value at time $i$, where closeness is defined as within 10 per cent of the maximum amplitude in flux within the light curve. The intensity of the blue dot correlates to $\epsilon = x(p) - x(i)$, the difference between the two points in the time series, divided by the maximum amplitude of the time series, i.e. the fractional 'closeness' of the two points in the time series. Regions of the time series that are close to repeating themselves appear as short horizontal lines. One such example is located within the black box aligned with the line in red, which corresponds to the regions plotted in red and blue in the time series of Fig.~\ref{fig:RRRx}.}
	\label{fig:closereturns}					
\end{figure*}

\begin{figure*}	
	\centering
	\includegraphics[width=0.9\textwidth]{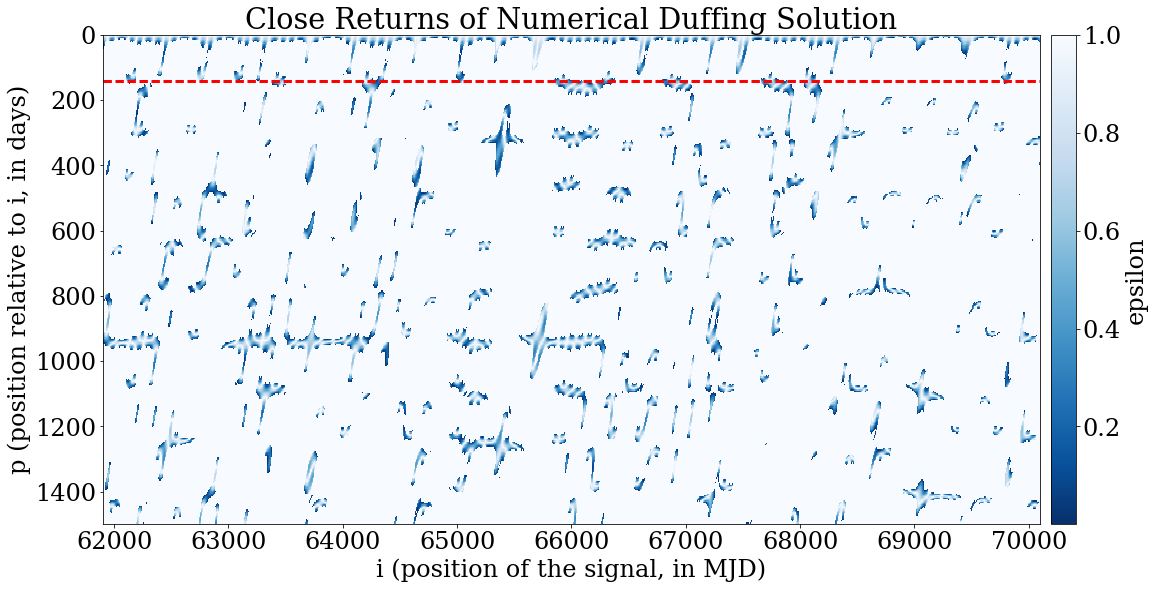}	
	\caption{The same close returns map as Fig.~\ref{fig:closereturns} for the reference Duffing solution. Note that the Duffing solution was created to be ten times the length of the 4U1705--44 time series. This close returns map is zoomed in to a shorter portion of the time series in order to highlight the regions of period-1 orbits. The dotted, red line corresponds to $p=140$ days, from which we can extract a 1D close returns plot for a single period, as in Fig.~\ref{fig:duffreturns}.}
	\label{fig:closereturnsDuff}					
\end{figure*}

A chaotic attractor is characterised by its periodic orbits, both stable and unstable. A topological analysis of the data can be performed once these periodic orbits are determined. Since the lowest-order driving period for 4U1705--44 is unknown and we would like to compare its value to the range of driving periods in the Duffing solutions, we use the method of close returns as utilised by \citet{Mindlin2} and \citet{Boyd94}, ideal for small, chaotic datasets to identify nearly closed orbits near unstable periodic orbits of low period. In Fig.~\ref{fig:closereturns} a blue point is plotted at $(i, p)$ when, for that time $i$, there is a point $p$ days later with flux that is 'close' to the flux value at time $i$. We define 'close' as within 10 per cent of the maximum amplitude of the time series, represented by the intensity of the colour map, $\epsilon$. The maximum amplitude is the difference between the maximum and minimum flux present within the time series. Regions of the time series that are close to repeating themselves appear as horizontal lines in blue. These correlate to sections in the light curve in which the time series comes close enough to an unstable periodic orbit such that it remains close for at least one period and is therefore within a region close to a periodic orbit. One example of a horizontal stretch is highlighted in Fig.~\ref{fig:closereturns}, corresponding to the regions in the light curve outlined in blue (portions of the light curve at position $i$) and red (the light curve at position $p$) in Fig.\ref{fig:RRRx}. We locate and extract all such regions from the light curve and find the lowest order orbits have periods between 110 and 200 days. For comparison, we computed the close returns map for the Duffing solution time series in Fig.~\ref{fig:closereturnsDuff}. The Duffing close returns map more distinctly shows regions of the time series that repeat itself, appearing as horizontal lines throughout, and without the noise that is present in the 4U1705--44 time series.

\begin{figure*}
	\centering
	\includegraphics[width=0.9\textwidth]{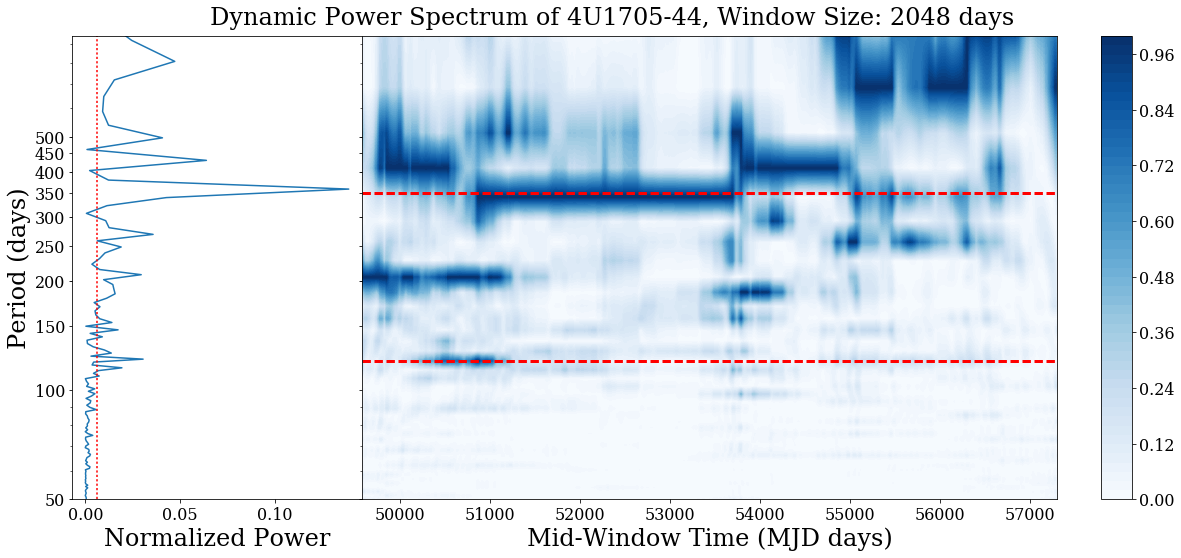}	
	\caption{Left panel: The total power spectrum of the 4U1705--44 light curve, normalised according to \protect\cite{Shimshoni}. 
	Vertical dotted red line corresponds to at least 99 per cent significance in power. Right panel: Dynamic Power Spectrum of 4U1705--44, 
	with a window size of 2048 days, normalised to unity. The edges of the time series are padded with Gaussian noise up to half the window size 
	in order to extract periodicities at the beginning and end of the dataset. We observe the lowest order periodicity occurs between 100 and 150 days, 
	especially early in the time series, with a strong period-3 presence developing later. The longer periods arising near the end of the time series are 
	likely a consequence of the persistent high state in the time series from about MJD 55000 and on. The dashed red lines correspond to periods of 
	120 and 350 days, for reference.}
	\label{fig:dynamicfft}					
\end{figure*}

We can constrain the lowest order period obtained from the close returns analysis with the dynamic power spectrum. Fig.~\ref{fig:dynamicfft}  shows the total power spectrum of the light curve of 4U1705--44 plotted against the period in the left pane. In order to determine the significance of the spectral peaks, we follow the \cite{Shimshoni} version of the Fisher test of significance in harmonic analysis \citep{Fisher}. The spectrum is normalised to the total power minus the first power term in the Fourier series. A vertical, dotted line is plotted in the total power spectrum in the left pane at a threshold value of $g=0.00646$ (in the notation of \cite{Shimshoni} and \cite{Fisher}) above which all peaks are considered significant at a 99 per cent confidence level or above.

Immediately, we observe all of the significant power in the spectrum is at periods longer than 100 days, corresponding to low frequencies, indicative of longer-term variability. The right pane is the dynamic power spectrum showing the power spectrum of a window size of 2048 days, centred on the middle day in the window as it evolves through the time series. That is, for a particular time in the time series, the power spectrum is plotted with the window centred at that particular time and the power at each frequency represented by the colour map. The ends are padded with random noise, out to half a window size, with the same standard deviation as the entire dataset in order to extract the periodicities in the ends of the light curve. 

\begin{figure}
	\centering
	\includegraphics[width=0.5\textwidth]{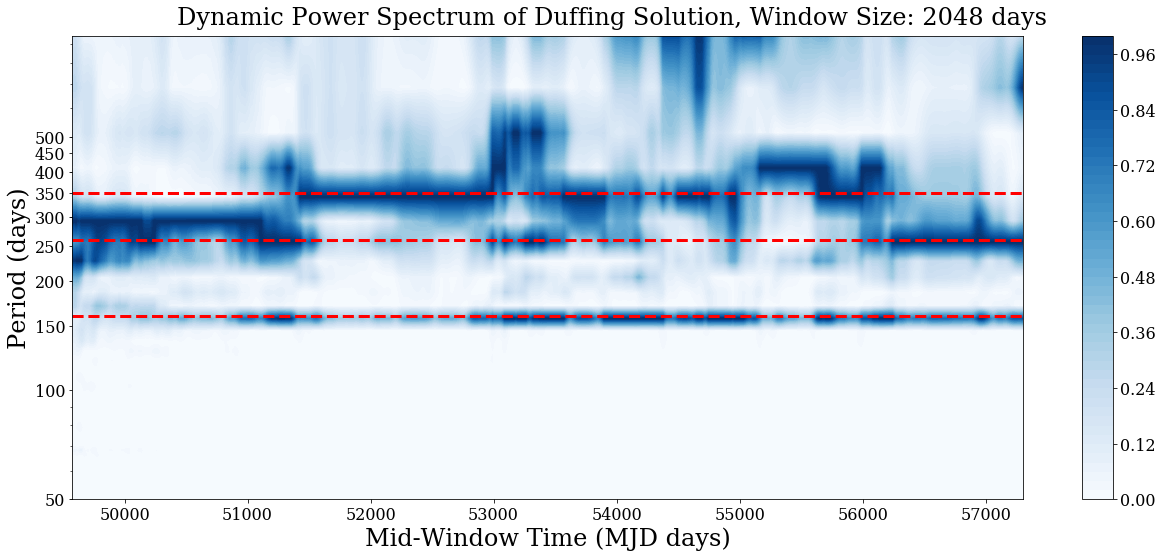}	
	\caption{Dynamic Power Spectrum of the reference Duffing solution time series, with a window size of 2048 days, normalised to unity. The edges of the time series are padded with Gaussian noise up to half the window size in order to extract periodicities at the beginning and end of the dataset. The dashed lines are located at periods of 160, 260, and 350 days, for reference.}
	\label{fig:dynamicFFTduff}					
\end{figure}

The lowest order period varies between 120 and 150 days, particularly in the first half of the time series, in agreement with the periods found in the close returns analysis. If we consider the total power spectrum in the left pane, we observe a cluster of peaks isolated right around 125 days. We also potentially see the appearance of modes of higher order periods, which could relate to period-2 and period-3 orbits in the data (twice and three times the lowest-order period). For a first-order period of 120 days, we would expect a period-3 to be at 360 days and we do in fact observe the strongest peak in the total power spectrum to occur near this value. The very long period near the end of the light curve is likely related to 4U1705--44 remaining in a high-state for hundreds of days, a deviation from its previous behaviour, in which longer wavelength Fourier modes on the order of the size of the window could be favoured. For comparison, we have also computed the dynamic power spectrum for the reference Duffing solution in Fig.~\ref{fig:dynamicFFTduff}. Most notably, we observe similar modal behaviour at similar frequencies to the 4U1705--44 power spectrum.

\begin{figure}
	\centering
	\includegraphics[width=0.5\textwidth]{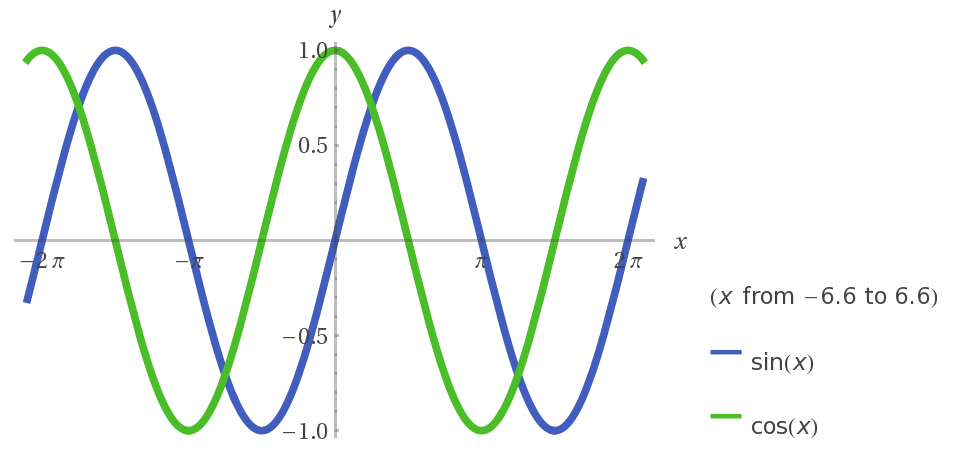}	
	\caption{Plot of $\sin(x)$ and $\cos(x)$, which are phase-shifted by $\pi/2$, a quarter of the lowest-order period of $\sin(x)$ of $2\pi$. If we were to apply a time-delay embedding on $\sin(x)$, we would recover its derivative, $\cos(x)$ using a time-delay of $\pi/2$. We seek to find the corresponding time-delay embedded in the 4U1705--44 data that would recover its phase space trajectory. The time-delay is related to the lowest-order period in the 4U1705--44 time series.}
	\label{fig:sine}					
\end{figure}

Finally, we can consider the periodic information we obtain from a time-delay embedding of the 4U1705--44 time series. A strange attractor is embedded in a $\geq 3$-dimensional space and topological and geometrical analyses in chaos theory are performed on the embedded $n$-dimensional phase space of the attractor. One can reconstruct a strange attractor from scalar data, $x(t)$, like a light curve, by constructing vectors $y(t)$ with $n$ components determined by the scalar $x(t)$. One such embedding is determined as the differential phase space projection, as in Fig.~\ref{fig:phasespace}, where $y(t)$ is merely the derivative of $x(t)$. Another is a time-delay embedding. This involves creating an $n$ vector by the map
\begin{equation}
\begin{split}
	x(t) \rightarrow & y(t) = (y_1(t), y_2(t), \dots, y_n(t)) \\
				& y_k(t) = x(t-\tau_k), \,\,\, k=1,2,\dots,n,
\end{split}	
\label{eq:timedelay}
\end{equation}
where $\tau_k$ are called the time delays. The time delays are typically evenly spaced multiples of $the$ time delay: $\tau_k = (k-1)\tau$. According to \citet{Takens}, we can reconstruct a mapping of the strange attractor from the time series via Eq.~\ref{eq:timedelay} by merely introducing a time-related shift. By adjusting the time delay component, we can determine the value at which it approaches the differential phase space embedding, which is related to the lowest order period of the time series. A simple example of a time-delay embedding and its relationship to its derivative is that of $sine$ and $cosine$. That is, $cosine$ is the derivative of $sine$, which when shifted by a quarter of its period, perfectly overlaps $sine$, as exemplified in Fig.~\ref{fig:sine}. The shift that overlaps $sine$ with its derivative is related to its lowest-order period. In a time-delay phase-space, this becomes apparent when $sine$ is plotted against its time-delayed self. As the delay coordinate is increased, the phase-space projection progresses from a diagonal line (for a time-delay of 1 unit) and expands, or unfolds, towards the image of its differential phase-space projection, which would be a perfect circle for $sine$. Time-delay embeddings are traditionally examined via visual inspection (for other examples see \citealt{Takens}; \citealt{Casdagli1992}; \citealt{Sauer94}).

\begin{figure}
	\centering
	\includegraphics[width=0.5\textwidth]{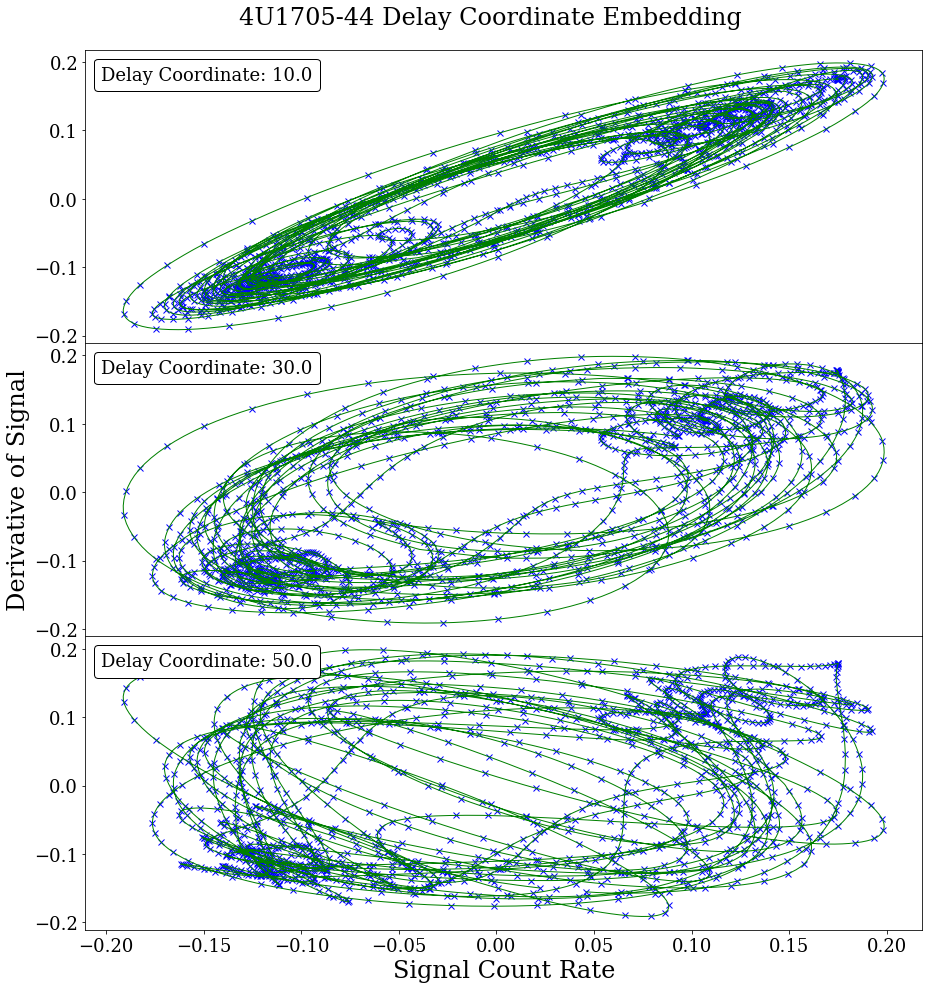}	
	\caption{Time-delay embedding of 4U1705--44, for time-delays of 10, 30, and 50 days. The point here is to recover an embedding that closely resembles the differential phase space embedding of the time series, Fig.~\ref{fig:phasespace}, which will be related to an intrinsic lowest order period in the data. The time-delay closest to the derivative phase-space embedding is traditionally determined via visual inspection and, in this case, occurs for time delays of 30-35 days, corresponding to driving periods of 120-140 days. This can be compared to the relationship between sine and cosine, which are derivatives of each other and resemble each other with an offset of some phase factor, or time-delay, related to the driving period.}
	\label{fig:delay}					
\end{figure}

From Fig.~\ref{fig:delay}, we look for the optimal unfolding with increasing time delay of 4U1705--44 towards its image in the differential phase space projection of Fig.~\ref{fig:phasespace}, which occurs roughly between 30 and 35 days. As the time delay is a quarter of the signal's period, the lowest-order period from this embedding is between 120 and 140 days. If we were to continue to increase the time-delay coordinate, we would see the phase space projection continue to expand away from the image of the differential phase space embedding and then come back towards it at the next underlying period (e.g towards the next large peak in the power spectrum). This is similar to encountering the successive peaks and zero-crossings of the autocorrelation function of the light curve. For our purposes, we are merely looking for a range of values relating to the lowest-order driving period consistent with the power spectrum and close returns analysis.

\section{Extracting Unstable Periodic Orbits}\label{sec:Extract}
\begin{figure}
\centering
\subfloat[]{
	\includegraphics[width=0.5\textwidth]{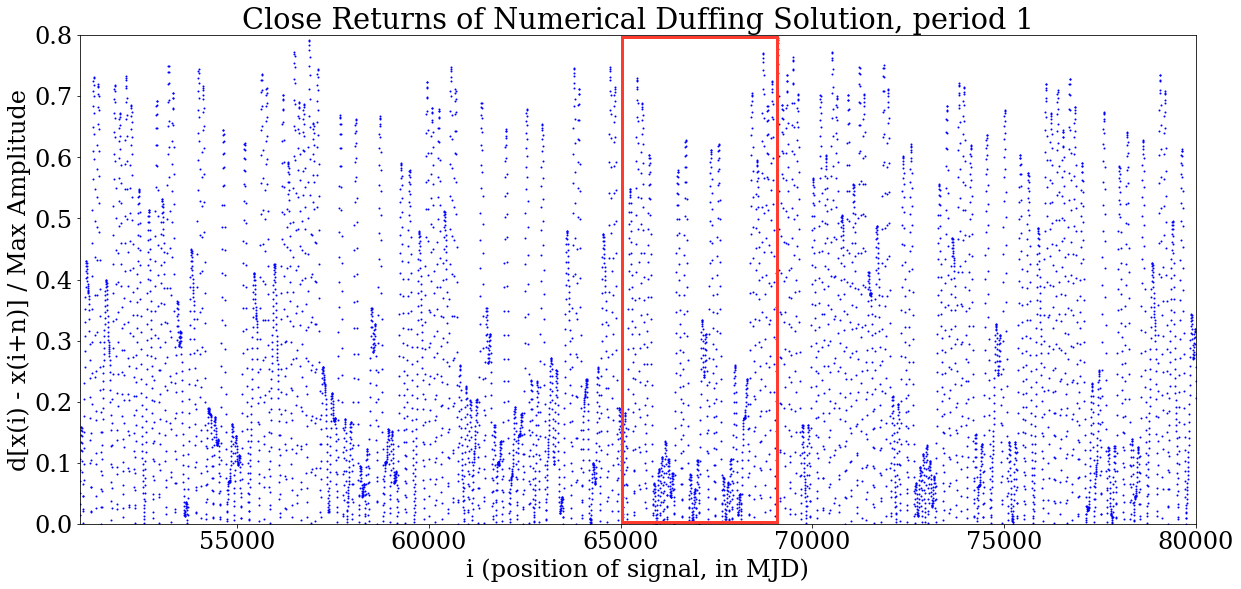}}
\qquad
\subfloat[]{
 	\includegraphics[width=.5\textwidth]{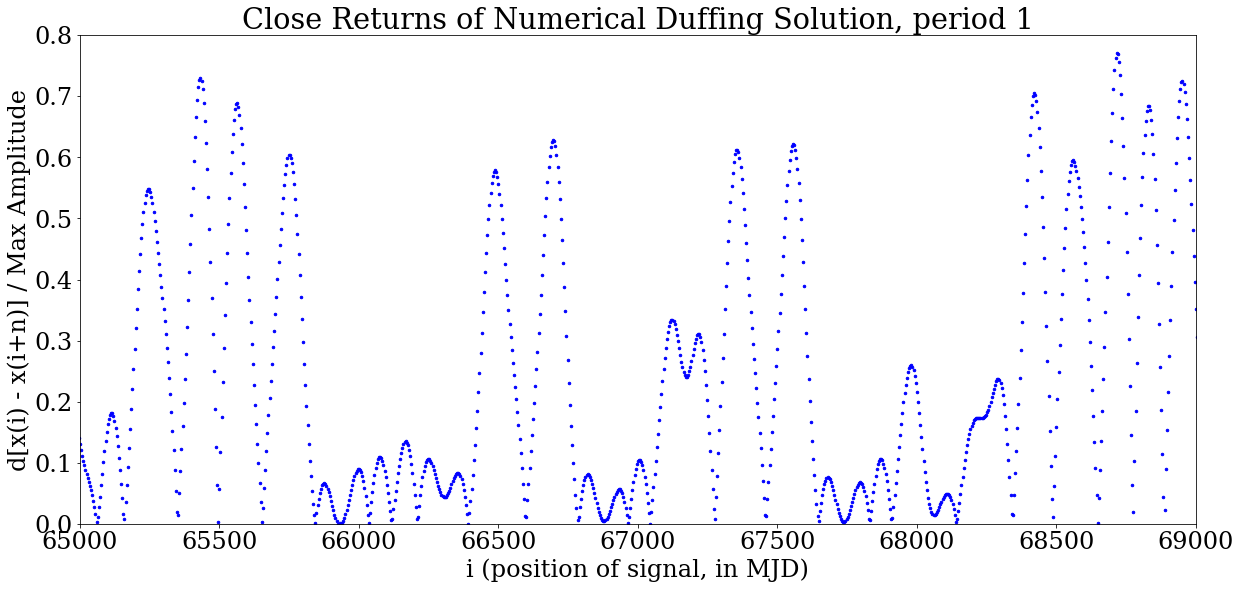}}
\caption{Relative distances between the points x(i) and x(i+n), $d[x(i)-x(i+n)]$ as a fraction of the total amplitude of the time series, plotted as a function of $i$, for a fixed period $n$. In this example, we are searching for period-1 orbits, which are of length 140 days; (a) is of a large sample of the time series, (b) is a zoomed in portion highlighted by the red box in (a) in which we find a few candidate period-1 orbits. We are hunting for regions in which the fractional difference between a point and one 140 days later is under 0.1 (or 10 per cent of the maximum flux amplitude) and persisting for at least one period (or $\geq$140 days). This shows us the location of potential unstable periodic orbits, identified as a region where the time series is close to repeating itself for at least one period. We can then locate the same regions within the Duffing time series (e.g. about days 66000 MJD), and extract and label these sub-sections of the time series as surrogate period-1 orbits.}
\label{fig:duffreturns}
\end{figure}

We have determined a range for the lowest-order period in the 4U1705--44 data. We have also produced numerical solutions of the Duffing oscillator. Its lowest order period is simply the driving period, $\omega$, from the Duffing equation, which we know to be 140 days for the solution plotted in Fig.~\ref{fig:duffphasespace}. Starting with the numerically generated Duffing time series, we can extract low-order, unstable periodic orbits from the time series by using a 1-D version of the close returns method. For a particular period we can compare each position in the time series to each successive period later, calculate the relative distance and, if this distance falls within a small value, $\epsilon$, and evolves within that neighbourhood for at least one period, then we have located an unstable periodic orbit \citep{Mindlin2}. This is equivalent to extracting the row located at $p=140$ days (highlighted by the red, dashed line) in Fig.~\ref{fig:closereturnsDuff} for the Duffing time series and plotting its position, $i$, against $\epsilon$. Such a determination is exemplified in Fig.~\ref{fig:duffreturns} for a period-1 orbit and is distinct from the full close returns mapping as it searches for periodicities of a specific period one at a time. Once we have located the region in the time series where a periodic orbit exists (e.g. a period-1 orbit is located about 55800 MJD identified by Fig.~\ref{fig:duffreturns}(b)), and do so for as many periods as we can discover in the time series, then we can save each of these periodic orbits in order to compute the relative rotation rates.

Using the method of close returns on the numerically generated Duffing solutions, we were able to extract unstable periodic orbits up to period-6 by varying the parameter $n$ in Fig.~\ref{fig:duffreturns}. That is, we were able to locate and extract 5 period-1, 4 period-2, 3 period-3, 4 period-4, 2 period-5, and 2 period-6 orbits.

Given that we determined the lowest order period in the 4U1705--44 data to be in the rough range of 120 to 170 days, we chose the Duffing period-1 length of 140 days for optimal comparison. Using the same method, we thereby reduced the close returns plot in Fig.~\ref{fig:closereturns} to a one-dimensional version as done in the Duffing case. As a result, we located regions of unstable periodic orbits and successfully extracted three orbits for each period-1, -2 and -3.

\section{Relative Rotation Rates}\label{sec:RRR}
A strange attractor can be characterised by the invariants that it possesses. The traditional classification of a strange attractor is by the determination of its dynamical and metric invariants, e.g. the Lyapunov exponents and various dimensions (Correlation or Minkowski). The other type of invariant is topological. Where the first two types are invariant only under coordinate transformations, the topological type is also invariant under control-parameter changes. For experimental conditions in which the control parameters experience perturbations the determination of the topological invariants is appropriate \citep{Solari1} for up to three-dimensional attractors. Furthermore, the topological approach can be applied to noisy systems and shorter datasets.

Topological invariants draw dependence on the periodic orbits existing in a strange attractor. The mechanisms that drive the behaviour of a strange attractor uniquely organize all the unstable periodic orbits embedded in that strange attractor. The 'stretching' mechanism is correlated to the sensitivity to initial conditions (nearby points essentially repel each other at an exponential rate), whilst the 'squeezing' mechanism is responsible for keeping the trajectory bounded in phase space. Identifying the organisation of the unstable periodic orbits can be used to identify these underlying mechanisms and thereby provide us with the 'fingerprints' on which the attractor is built \citep{Mindlin1}. The underlying structure described by these mechanisms is completely responsible for the organisation of all periodic orbits in a flow. This becomes a powerful mathematical tool because if we can make a statement about the equivalence of the underlying mechanisms driving two systems, then we can qualitatively compare these two systems. Most notably, in the case of the Duffing oscillator and 4U1705--44, we can look to the Duffing equation to garner insights into the long-term behaviour of 4U1705--44.

One of the topological invariants introduced by \citet{Solari2}, intended to describe periodically driven two-dimensional dynamical systems, such as nonlinear oscillators, is called the Relative Rotation Rates (RRR). 
Generally speaking, RRR describe how one periodic orbit rotates around another; or, more specifically, the average value, per period, of this rotation rate.

The RRR can be computed only after the orbits have been embedded in a three-dimensional space \citep{Gilmore}. 
We chose a three-dimensional differential phase space embedding \citep{Mindlin2}:
\begin{equation}\label{eq:embed}
	\begin{split}
	x(i) \rightarrow y(i) &= \{x(i), dx(i)/dt, d^2x(i)/dt^2\} \\
	&\approxeq \{ x(i), \, x(i+1) - x(i-1), \\
	    & x(i+1) - 2x(i) + x(i-1) \}
	\end{split}
\end{equation}

Let A be an orbit of period $p_A$ which has intersections $(a_1, a_2, \ldots, a_{p_A})$ with a Poincar$\acute{e}$ section $t=const$, and similarly for orbit B. The relative rotation rate $R_{ij}(A,B)$ of A around B for the initial conditions $(i,j)$ is defined as:
\begin{equation}\label{eq:RRR1}
	R_{ij}(A,B) = \frac{1}{2\pi p_A p_B} \oint \frac{ \boldsymbol{n} \cdot (\boldsymbol{\Delta r} \times d\boldsymbol{\Delta r})}{\boldsymbol{\Delta r} \cdot \boldsymbol{\Delta r}}
\end{equation}
where $\boldsymbol{\Delta r} = [x_B(t) - x_A(t), y_B(t)-y_A(t)]$ is the difference vector between points on the two orbits, $\boldsymbol{n}$ is the unit vector orthogonal to the plane spanned by $\boldsymbol{\Delta r}$ and $d \boldsymbol{\Delta r}$, and the integral extends over $p_A \times p_B$ periods. The initial conditions are the points $a_i, b_j$ on the Poincar$\acute{e}$ section and the resulting $R_{ij}(A,B)$ is a single number represented as a fraction for the orbits of A and B. All of the relative rotation rates for a system can be assembled into a table, or 'intertwining matrix' \citep{Solari1}.

\begin{figure}
	\centering
	\includegraphics[width=0.48\textwidth]{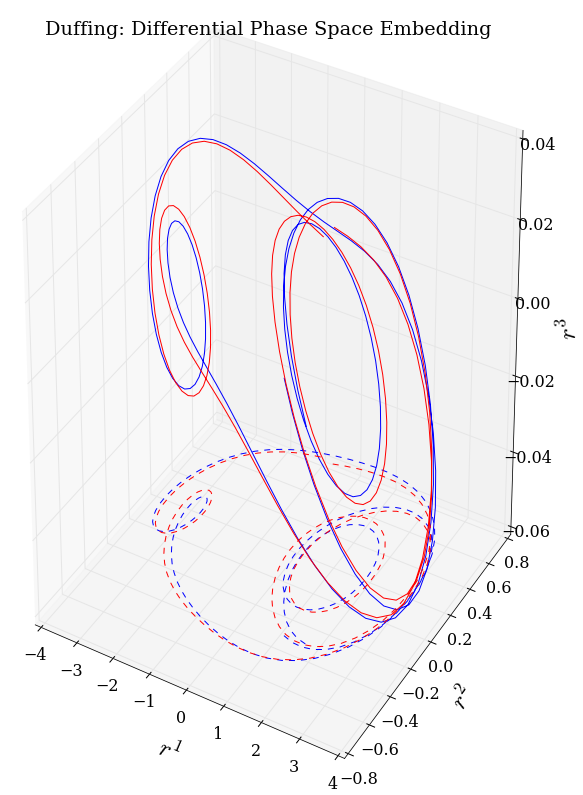}
	\caption{A 3D phase space embedding and 2D projection of a pair of extracted unstable periodic orbits of period 4 from the numerical Duffing solutions over a period 4$T$.}
	\label{fig:RRR}
\end{figure}

An example of the 3D embedding of an extracted period-4 orbit plotted against another period-4 orbit in the numerical Duffing time series is plotted in Fig.~\ref{fig:RRR}. To compute the RRR, we located the intersection of each extracted periodic orbit with the same Poincar$\acute{e}$ section. Next, we connected each pair of points in the two orbits by a directed line segment. This line segment will evolve in time under the flow and will have undergone an integer number of full rotations ($2\pi$ radians) in the plane perpendicular to the flow in $p_A \times p_B$ periods. The relative rotation rates can be computed in four equivalent ways as described by \cite{Solari1}. The first is the mathematical description as defined in Eq.~\ref{eq:RRR1} as an integral for continuous datasets. The line integral essentially describes the length of the rotating difference vector per period through phase space. The line integral can be described using graphical means, much like the trajectory of a time series can be plotted and the area under the curve can be computed by eye. One of these other methods, a graphical approach and therefore very straightforward for our means, is as follows: 

Whenever $\Delta r^2 = 0$ and $\Delta r^1 \geq 0$, define $\sigma(t)$:

\begin{equation}
	\sigma(t) =
	\begin{cases} 
      		+1 & d\Delta r^2 / dt > 0 \\
      		-1 & d\Delta r^2 / dt < 0 
   	\end{cases}
\end{equation}
Then
\begin{equation}\label{eq:RRR2}
	R_{ij}(A,B) = \frac{1}{p_A p_B} \sum_{0 \leq t \leq T_{p_Ap_B}} \sigma(t)
\end{equation}

\begin{figure}
	\centering
	\includegraphics[width=0.48\textwidth]{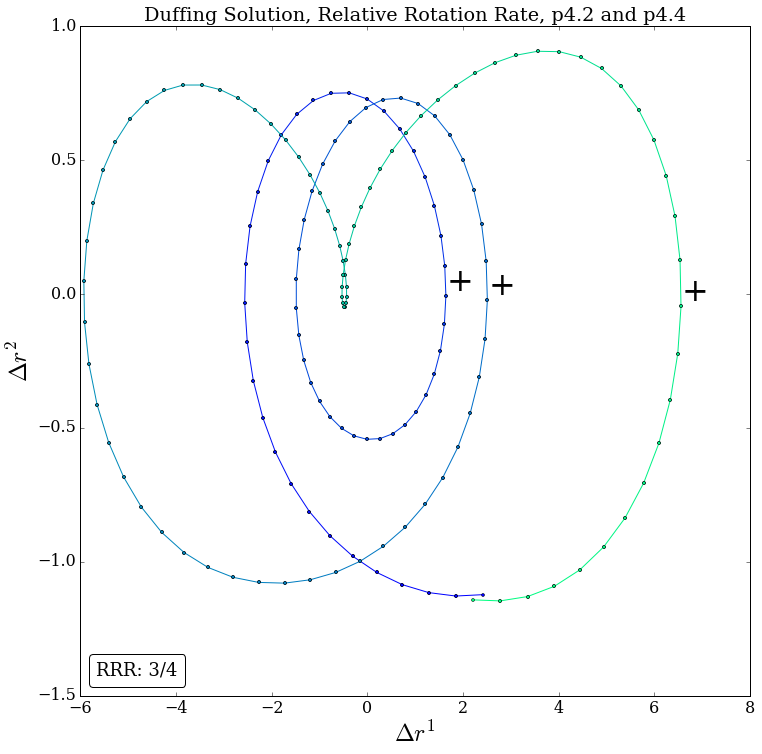}
	\caption{Computation of RRR for two period-4 orbits of Fig.~\ref{fig:RRR}. Each time the difference $\Delta r$ crosses the half line, $\Delta r^1$, crossing direction is positive ($\sigma = +1$); oppositely for negative. This is shown over a period of 4T (rather than $p_A \times p_B$). }
	\label{fig:RRR1}
\end{figure}

In summary, we sum the number of times the difference vector crosses the positive x-axis from above or below per total period. Fig.~\ref{fig:RRR1} shows each time the difference $\Delta r$ crosses the half line $\Delta r^2 = 0, \, \Delta r^1 > 0$, whereby the crossing direction is counted positive ($\sigma = +1$) if $d\Delta r^2 /dt > 0$ and negative if $d\Delta r^2 /dt < 0$. For the period-4 compared against a different period-4, we find the RRR is 3/4 for the Duffing solution.

\begin{figure}
  	\centering
  	\subfloat{\includegraphics[width=.48\textwidth]{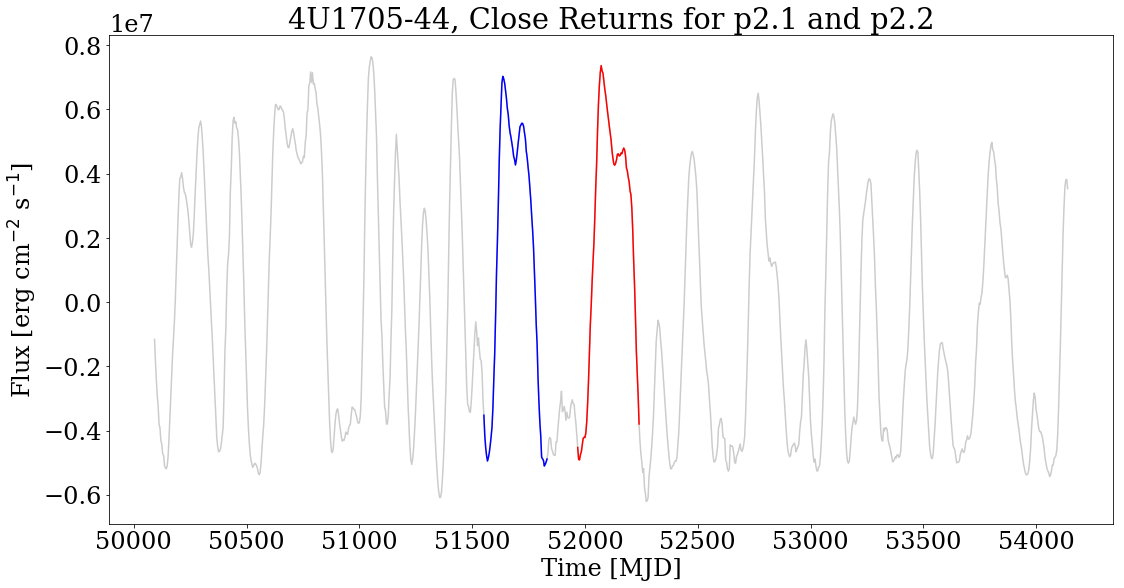}}\\
	\centering
  	\subfloat{\includegraphics[width=.48\textwidth]{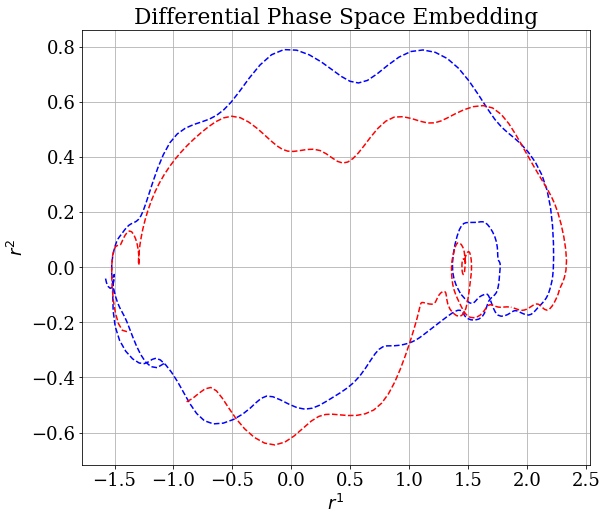}}
	\caption{Bottom Panel: A 2D phase space embedding (akin to the projected panel in Fig.~\ref{fig:RRR} and normalised) of a pair of extracted unstable periodic orbits of period 2 from the 4U1705--44 embedded time series, highlighted within the original time series of 4U1705--44 (Top Panel). These two unstable periodic orbits of period 2 were identified using the Close Returns method, highlighted in Fig.~\ref{fig:closereturns}.}
	\label{fig:RRRx}
\end{figure}

\begin{figure}
	\centering
	\includegraphics[width=0.48\textwidth]{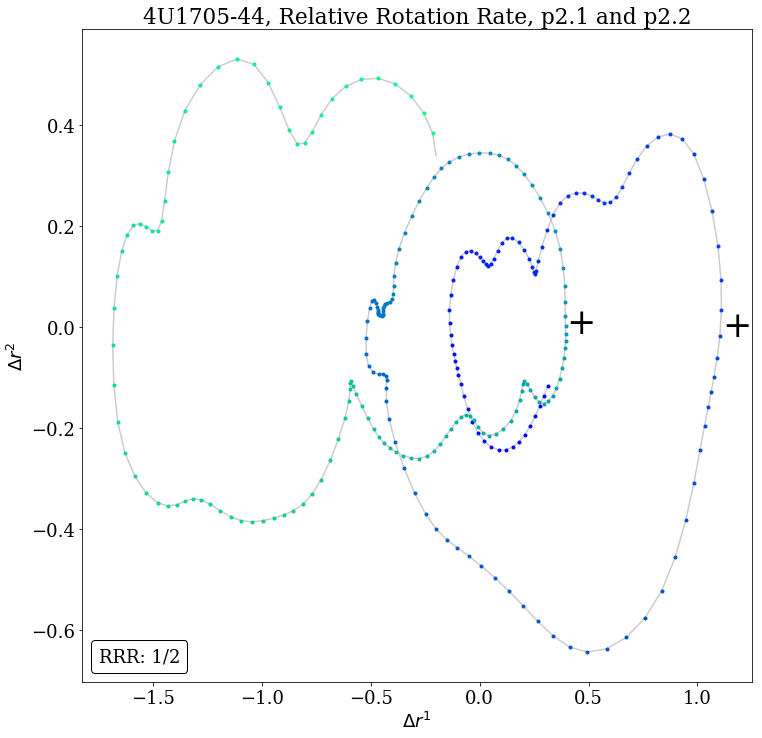}
	\caption{Computation of RRR for two period-2 orbits of Fig.~\ref{fig:RRRx}. Each time the difference $\Delta r$ crosses the half line, $\Delta r^1$, the crossing direction is positive ($\sigma = +1$); oppositely for negative. We observe similar behaviour to the period-4 Duffing orbits.}
	\label{fig:RRR1x}
\end{figure}

Following the same procedure with all extracted unstable periodic orbits, we collect the RRR for both 4U1705--44 and the numerically generated Duffing time series into their respective intertwining matrices, where we have given each extracted orbit a different identifying name according to its behaviour in phase space. An example of extracted surrogate, unstable period-2 orbits (akin to Fig.~\ref{fig:RRR}) from the 4U1705--44 light curve in a 2D phase-space embedding is plotted in Fig.~\ref{fig:RRRx} and the graphical computation of the RRR for the same period-2 orbits is shown in Fig.~\ref{fig:RRR1x}.

In computing the RRR, we noted that the time at which we began collecting data on 4U1705--44 is arbitrary to its mathematical representation. We therefore assumed that the initial point of the first orbit appearing in the time series intersected a Poincar$\acute{e}$ section and then chose the initial point of the second orbit such that it intersected the same Poincar$\acute{e}$ section. We then varied the initial point of the first orbit within the range that it was still close to an unstable periodic orbit, matched the second orbit to the new Poincar$\acute{e}$ section, and computed the RRR again. We found no changes in the computation of the RRR for each pair of orbits when varying the initial point. This was important for self-verification as the numerical Duffing time series is much smoother than the 4U1705--44 time series.
Up through period-3 (the longest period extracted from the 4U1705--44 data) the matrices are identical. Given that these matrices are suggestively identical, it is therefore possible that the flows are equivalent. This method is essentially a means to falsify a set of differential equations and we have shown that the Duffing oscillator, and its family of differential equations, cannot yet be ruled out as related to the equations of motion generating the 4U1705--44 light curve. As such, if the RRR do continue to agree, then the underlying template of the Duffing oscillator can serve as a geometric model for the dynamics which generate the chaotic time series of the 4U1705--44 system. 

\begin{table}
\begin{tabular}{ p{0.5cm}||p{0.5cm}p{0.5cm}p{0.5cm}p{0.5cm}p{0.5cm}p{1cm}p{1cm} }
 \hline
 \multicolumn{8}{c}{4U1705--44 Relative Rotation Rates} \\
 \hline
 & 1.1 & 1.2 & 1.3 & 2.1 & 2.2 & 3.1 & 3.2 \\
 & $x\gamma$ & $x\gamma$ & $x\gamma$ & $yy\alpha$ & $yy\beta$ & $xyx\alpha$ & $xyx\beta$ \\
 \hline
1.1   & 0   & 1 & 1 & 1 & 1 & 2/3 & 2/3 \\
 1.2 &  & 0 & 1 & 1 & 1 & 2/3 & 2/3\\
1.3 & & & 0 & 1 & 1 & 2/3 & 2/3\\
2.1 & & & & 0 & 1/2 & 2/3 & 2/3 \\
 2.2 & & & & &0 & 2/3 & 2/3 \\
 3.1 & & & & & & 0 & 2/3\\
 3.2 & & & & & & & 0\\
 \hline
\end{tabular}
\caption{Orbits are labeled as (p.n), i.e. the $n^{th}$ orbit of period $p$; $x$ and $y$ signify the orbit's presence in each lower- or upper- well, $\gamma$ signifies symmetric orbits, and $\alpha, \beta$ asymmetric orbits. Only distinct orbits are included in the matrix.}
\end{table}

\begin{table}
\begin{tabular}{ p{0.5cm}||p{0.5cm}p{0.5cm}p{0.5cm}p{0.5cm}p{0.5cm}p{1cm}p{1cm} }
 \hline
 \multicolumn{8}{c}{Duffing Relative Rotation Rates} \\
 \hline
 & 1.1 & 1.2 & 1.5 & 2.1 & 2.2 & 3.1 & 3.2 \\
 & $x\gamma$ & $x\gamma$ & $x\gamma$ & $yy\alpha$ & $yy\beta$ & $xyx\alpha$ & $xyx\beta$ \\
 \hline
1.1   & 0   & 1 & 1 & 1 & 1 & 2/3 & 2/3 \\
1.2 &  & 0 & 1 & 1 & 1 & 2/3 & 2/3 \\
1.5 & & & 0 & 1 & 1 & 2/3 & 2/3 \\
2.1 & & & & 0 & 1/2 & 2/3 & 2/3 \\
2.2 & & & & &0 & 2/3 & 2/3 \\
3.1 & & & & & & 0 & 2/3\\
3.2 & & & & & & & 0 \\
 \hline
\end{tabular}
\caption{RRR of numerically generated Duffing time series with the same identifying naming convention of orbits as 4U1705--44 RRRs.
		Orbits of the same type as in 4U1705--44 are included here.}
\end{table}

\section{Discussion}\label{sec:Conclusions}
In the previous sections we have presented evidence that the time evolution of X-ray binary 4U1705--44 shares key features with the nonlinear Duffing equation in a chaotic regime. The mechanism in 4U1705--44 and other X-ray binaries showing super-orbital variability is not well understood. Our results suggest that a double-welled, non-linear oscillator is a strong candidate to describe these systems. Specifically, we have found that the low-order driving period lies near 125 days, as is seen in the power spectra, time-delay embedding and close returns analysis. Similarly, the driving frequency ($\omega$) of the reference Duffing solution has a similar driving period of around 140 days. 

The relative rotation rates analysis suggests that 4U1705--44 and the Duffing equation share the same underlying template. This means we can look to the Duffing equation to provide insight into the behaviour of the 4U1705--44 system. For example, we can consider the five parameters of the Duffing equation and how each might relate to the dominant physical processes in the neutron star binary system. 

\begin{figure}
	\centering
	\includegraphics[width=0.4\textwidth]{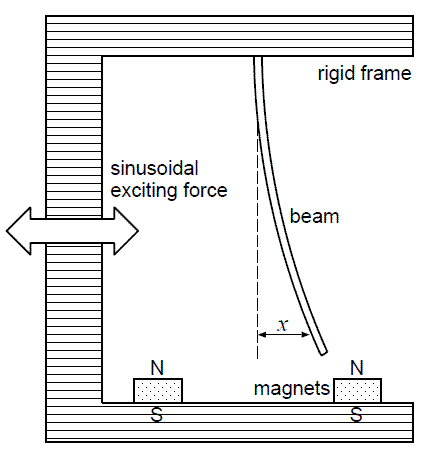}
	\caption{Toy model of the Duffing oscillator \citep{DuffFigure} whereby the position $x(t)$ of the metal beam oscillates chaotically with time between the two magnets, all attached to a sinusoidally driven rigid frame.}
	\label{fig:toymodel}
\end{figure}

The Duffing oscillator can be described by a flexible metal beam attached to a rigid, oscillating frame whereby the beam moves between two magnets (Fig.~\ref{fig:toymodel}). Returning to the Duffing equation, Eq.~\ref{eq:duff}, we see the flexibility (or stiffness) of the metal beam is described by $\alpha$, while $\beta$ describes the non-linearity of the metal beam material, much like a position-varying spring constant. The rigid frame is driven harmonically, described by frequency $\omega$, where the driving has an amplitude of $\gamma$. The rigid frame drives both the metal beam and the two magnets on either side, which means that the driving frequency correlates the motion of the metal beam to that of the magnets. Furthermore, strong enough magnets would force the metal beam to oscillate at the driving frequency.

With a picture of a desk-sized oscillator in mind, we can hypothesise the relationship of each component of the toy model to physical mechanisms in the neutron star binary system. Perhaps the most important is the source of the driving frequency. We know that the observed period in the time series of 4U1705--44 is over a hundred days, most likely occurring at approximately 120 days, and is therefore substantially longer than the probable orbital period of 1-10 hours \citep{Homan}. We therefore should consider precessional periods that might be present in the system, which would explain longer, super-orbital periods.

\cite{Postnov} found compelling evidence that there exists variable neutron star free precession in Hercules X-1 relating to its observed 35-day cycle. Free precession of a body that is not perfectly spherical results in an angle between a point on the body's surface and the total angular momentum vector and thus the location of the magnetic poles. The ellipticity of the neutron star body dictates the length of the precession period \citep{Landau}. For Her X-1, the neutron star non-sphericity should be $\epsilon \approx 10^{-6}$ resulting in a 35-day free precession period. In order to obtain a period of 120 days, as in the 4U1705--44 system, the non-sphericity would be approximately a third of Her X-1, if we assume the same parameters. That is, the 4U1705--44 neutron star would be more spherical than that of Her X-1.

For Her X-1, the time series is strongly periodic about the 35-day period. This suggests that the strength of the driving force is much higher in Her X-1 than in 4U1705--44. That is, if $\gamma$ determines the strength of the magnets in the Duffing toy model and thus the strength of the driving force, then the strength of the neutron star magnetic field, where the magnetic poles are misaligned, could slave the accretion disc to the precession period in Her X-1. In the case of 4U1705--44, where we observe more chaotic behaviour, the neutron star magnetic field might therefore be weaker, allowing for other parameters in the system to compete with the free precession period. A cyclotron line is present in the spectra of Her X-1, corresponding to a magnetic field of approximately $10^{12}$ G \citep{Shapiro}. In contrast, the non-detection of a cyclotron line by \cite{Fiocchi}, \cite{Piraino}, and \cite{Lin} and the estimated $10^{8}$ G magnetic field \citep{Lin} of the 4U1705--44 neutron star confirms our theory that 4U1705--44 has a weaker magnetic field than Her X-1.

A weaker magnetic field in 4U1705--44 allows for other physical mechanisms to compete with the free precession period of the neutron star driving the variability. For example, a disc-magnetosphere interaction, also considered by \cite{Postnov}, occurs when a magnetised neutron star is surrounded by a thin diamagnetic accretion disc. The magnetic field of the neutron star will exert a torque on the accretion disc, driving it out of the equatorial plane \citep{Lai}. As a result, the spin axis of a freely precessing neutron star in a binary system can remain tilted with respect to the orbital angular momentum vector thereby maintaining a tilt in the accretion disc out of the equatorial plane which can then precess. 

The companion to 4U1705--44 is a low-mass, dwarf star \citep{Homan}, a class of stars which are known to exhibit powerful flares on timescales from minutes to hours \citep{Hilton}. If the companion star is variable with a high-frequency of flare activity then it could have a significant impact on the observed variability.

We therefore have a cocktail of strong effects potentially producing the observed chaotic behaviour: free precession of the neutron star providing the clock of a 125 day driving period with the neutron star magnetic field modulating its strength; the disc-magnetosphere interaction dictating the non-linearity of the accretion disc; the variation in the companion star damping the supply of material in the accretion disc as well as the well-known magneto-rotational instability responsible for the viscosity mechanism in the accretion disc dictating the 'stiffness' of the oscillator.

There are other possible mechanisms to explain the long-term variability. For example, Lense-Thirring precession, which is a general relativistic frame-dragging effect due to the spin of a central massive object, has been examined for various parameters of low-mass neutron star binaries by \cite{Morsink} and is found to relate to the observed QPOs in such systems. However, the computed precession frequencies (20-35 Hz) are still many orders of magnitude too short to be related to the driving period of over a hundred days.

Another promising source for the long-term driving period could be due to an irradiation-driven warped accretion disc. According to \cite{Pringle96}, an accretion disc that is initially flat is unstable to radiation-driven warping due to a central illuminating source. The resulting warp can then precess on the order of hundreds of days. This effect has been studied using smooth-particle-hydrodynamics (SPH) simulations of low-mass X-ray binaries by  \cite{Ogilvie2}, \cite{Foulkes1}, and \cite{Foulkes2}. The warp that develops due to a central source is predicted to occur in the outer portions of the accretion disc, as opposed to the central flow warp that develops due to the Lense-Thirring effect, and therefore is a more likely source of a longer period driving force. 

Any physical model chosen should also be in agreement with the observed spectral state transitions that occur in 4U1705--44 and the corresponding physical explanations. For example, \cite{Piraino2016} used a double Comptonisation model to describe the hard and soft state transitions of 4U1705--44 observed with \textit{BeppoSAX}. Their model suggests an alternating dominance of soft Comptonisation from a hot plasma surrounding the neutron star in the soft state and a hard Comptonisation arising from the disc in the hard state. 

We can verify the template identification of 4U1705--44 by computing the intertwining matrices and linking numbers for other numerically generated Duffing time series as well as from other external computations of the Duffing oscillator \citep{Tufillaro}. In a future work, we will extend this analysis to other X-ray binaries to investigate the presence of non-linearity and chaos in their light curves and whether the underlying template of the Duffing equation, or its family of differential equations, continues to be topologically equivalent to those generating the observed variability of other accretion powered systems. Already, the intertwining matrix for another chaotic system, the Belousov-Zhabotinsky chemical reaction \citep{Mindlin2}, is not identical to 4U1705--44 and differs by a non-integer (thus, we can predict that the flows and the equations that describe them are inequivalent). We also intend, in a future publication, to compare our analysis with the results of other methods aimed at exploring the source of the long periods in X-ray binaries such as simulations of accretion disc systems subjected to a central irradiating source or magnetically driven instabilities.

\section*{Acknowledgements}
This material is partially based upon work supported by the National Aeronautics and Space Administration under Grant Number NNX16AT15H issued through the NASA Education Minority University Research Education Project (MUREP) through the NASA Harriett G. Jenkins Graduate Fellowship activity. This research made use of data and software provided by the High Energy Astrophysics Science Archive Research Center (HEASARC), which is a service of the Astrophysics Science Division at NASA/GSFC and the High Energy Astrophysics Division of the Smithsonian Astrophysical Observatory. We also thank Dr. Robert Gilmore, author of \textit{The Topology of Chaos: Alice in Stretch and Squeezeland} and \textit{The Symmetry of Chaos} for helpful discussions on the topics of chaos, topology, and applications to real datasets.

\bibliography{refs}

\begin{thebibliography}{}
\makeatletter
\relax
\def\mn@urlcharsother{\let\do\@makeother \do\$\do\&\do\#\do\^\do\_\do\%\do\~}
\def\mn@doi{\begingroup\mn@urlcharsother \@ifnextchar [ {\mn@doi@}
  {\mn@doi@[]}}
\def\mn@doi@[#1]#2{\def\@tempa{#1}\ifx\@tempa\@empty \href
  {http://dx.doi.org/#2} {doi:#2}\else \href {http://dx.doi.org/#2} {#1}\fi
  \endgroup}
\def\mn@eprint#1#2{\mn@eprint@#1:#2::\@nil}
\def\mn@eprint@arXiv#1{\href {http://arxiv.org/abs/#1} {{\tt arXiv:#1}}}
\def\mn@eprint@dblp#1{\href {http://dblp.uni-trier.de/rec/bibtex/#1.xml}
  {dblp:#1}}
\def\mn@eprint@#1:#2:#3:#4\@nil{\def\@tempa {#1}\def\@tempb {#2}\def\@tempc
  {#3}\ifx \@tempc \@empty \let \@tempc \@tempb \let \@tempb \@tempa \fi \ifx
  \@tempb \@empty \def\@tempb {arXiv}\fi \@ifundefined
  {mn@eprint@\@tempb}{\@tempb:\@tempc}{\expandafter \expandafter \csname
  mn@eprint@\@tempb\endcsname \expandafter{\@tempc}}}

\bibitem[\protect\citeauthoryear{Anishchenko, Astakhov, Neiman, Vadivasova  \&
  Schimansky-Geier}{Anishchenko et~al.}{2007}]{Anishchenko}
Anishchenko V.~S.,  Astakhov V.,  Neiman A.,  Vadivasova T.,   Schimansky-Geier
  L.,  2007, Nonlinear Dynamics of Chaotic and Stochastic Systems: Tutorial and
  Modern Developments (Springer Series in Synergetics).
Springer

\bibitem[\protect\citeauthoryear{{Barret} \& {Olive}}{{Barret} \&
  {Olive}}{2002}]{Barret}
{Barret} D.,  {Olive} J.-F.,  2002, \mn@doi [\apj] {10.1086/341626}, \href
  {http://adsabs.harvard.edu/abs/2002ApJ...576..391B} {576, 391}

\bibitem[\protect\citeauthoryear{{Birman} \& {Williams}}{{Birman} \&
  {Williams}}{1983a}]{Birman2}
{Birman} J.~S.,  {Williams} R.~F.,  1983a, Contemp. Math., 20, 1

\bibitem[\protect\citeauthoryear{{Birman} \& {Williams}}{{Birman} \&
  {Williams}}{1983b}]{Birman1}
{Birman} J.~S.,  {Williams} R.~F.,  1983b, \mn@doi [Topology]
  {10.1016/0040-9383(83)90045-9}, 22, 47

\bibitem[\protect\citeauthoryear{{Boyd} \& {Smale}}{{Boyd} \&
  {Smale}}{2004}]{Boyd2004}
{Boyd} P.~T.,  {Smale} A.~P.,  2004, \mn@doi [\apj] {10.1086/421078}, \href
  {http://adsabs.harvard.edu/abs/2004ApJ...612.1006B} {612, 1006}

\bibitem[\protect\citeauthoryear{{Boyd}, {Mindlin}, {Gilmore}  \&
  {Solari}}{{Boyd} et~al.}{1994}]{Boyd94}
{Boyd} P.~T.,  {Mindlin} G.~B.,  {Gilmore} R.,   {Solari} H.~G.,  1994, \mn@doi
  [\apj] {10.1086/174496}, \href
  {http://adsabs.harvard.edu/abs/1994ApJ...431..425B} {431, 425}

\bibitem[\protect\citeauthoryear{{Boyd}, {Smale}, {Homan}, {Jonker}, {van der
  Klis}  \& {Kuulkers}}{{Boyd} et~al.}{2000}]{Boyd2000}
{Boyd} P.~T.,  {Smale} A.~P.,  {Homan} J.,  {Jonker} P.~G.,  {van der Klis} M.,
    {Kuulkers} E.,  2000, \mn@doi [\apjl] {10.1086/312931}, \href
  {http://adsabs.harvard.edu/abs/2000ApJ...542L.127B} {542, L127}

\bibitem[\protect\citeauthoryear{{Bradt}, {Rothschild}  \& {Swank}}{{Bradt}
  et~al.}{1993}]{Bradt}
{Bradt} H.~V.,  {Rothschild} R.~E.,   {Swank} J.~H.,  1993, \aaps, \href
  {http://adsabs.harvard.edu/abs/1993A%26AS...97..355B} {97, 355}

\bibitem[\protect\citeauthoryear{Casdagli}{Casdagli}{1992}]{Casdagli1992}
Casdagli M.,  1992, Journal of the Royal Statistical Society. Series B
  (Methodological), 54, 303

\bibitem[\protect\citeauthoryear{Casdagli}{Casdagli}{1997}]{Casdagli1997}
Casdagli M.,  1997, \mn@doi [Physica D: Nonlinear Phenomena]
  {https://doi.org/10.1016/S0167-2789(97)82003-9}, 108, 12

\bibitem[\protect\citeauthoryear{{Clarkson}, {Charles}, {Coe}, {Laycock},
  {Tout}  \& {Wilson}}{{Clarkson} et~al.}{2003a}]{Clarkson1}
{Clarkson} W.~I.,  {Charles} P.~A.,  {Coe} M.~J.,  {Laycock} S.,  {Tout} M.~D.,
    {Wilson} C.~A.,  2003a, \mn@doi [\mnras]
  {10.1046/j.1365-8711.2003.06176.x}, \href
  {http://adsabs.harvard.edu/abs/2003MNRAS.339..447C} {339, 447}

\bibitem[\protect\citeauthoryear{{Clarkson}, {Charles}, {Coe}  \&
  {Laycock}}{{Clarkson} et~al.}{2003b}]{Clarkson2}
{Clarkson} W.~I.,  {Charles} P.~A.,  {Coe} M.~J.,   {Laycock} S.,  2003b,
  \mn@doi [\mnras] {10.1046/j.1365-8711.2003.06761.x}, \href
  {http://adsabs.harvard.edu/abs/2003MNRAS.343.1213C} {343, 1213}

\bibitem[\protect\citeauthoryear{{Collins} \& {Scher}}{{Collins} \&
  {Scher}}{2002}]{Collins}
{Collins} G.~W.,  {Scher} R.~W.,  2002, \mn@doi [\mnras]
  {10.1046/j.1365-8711.2002.05844.x}, \href
  {http://adsabs.harvard.edu/abs/2002MNRAS.336.1011C} {336, 1011}

\bibitem[\protect\citeauthoryear{Collis, White  \& Hammond}{Collis
  et~al.}{1998}]{Collis}
Collis W.,  White P.,   Hammond J.,  1998, \mn@doi [Mechanical Systems and
  Signal Processing] {https://doi.org/10.1006/mssp.1997.0145}, 12, 375

\bibitem[\protect\citeauthoryear{{D'A{\`i}} et~al.,}{{D'A{\`i}}
  et~al.}{2010}]{D'Ai}
{D'A{\`i}} A.,  et~al., 2010, \mn@doi [\aap] {10.1051/0004-6361/200913758},
  \href {http://adsabs.harvard.edu/abs/2010A%26A...516A..36D} {516, A36}

\bibitem[\protect\citeauthoryear{Diks, van Houwelingen, Takens  \&
  DeGoede}{Diks et~al.}{1995}]{Diks}
Diks C.,  van Houwelingen J.,  Takens F.,   DeGoede J.,  1995, \mn@doi [Physics
  Letters A] {https://doi.org/10.1016/0375-9601(95)00239-Y}, 201, 221

\bibitem[\protect\citeauthoryear{{Durant}, {Cornelisse}, {Remillard}  \&
  {Levine}}{{Durant} et~al.}{2010}]{Durant}
{Durant} M.,  {Cornelisse} R.,  {Remillard} R.,   {Levine} A.,  2010, \mn@doi
  [\mnras] {10.1111/j.1365-2966.2009.15644.x}, \href
  {http://adsabs.harvard.edu/abs/2010MNRAS.401..355D} {401, 355}

\bibitem[\protect\citeauthoryear{Eckmann, Kamphorst  \& Ruelle}{Eckmann
  et~al.}{1987}]{Eckmann}
Eckmann J.-P.,  Kamphorst S.~O.,   Ruelle D.,  1987, EPL (Europhysics Letters),
  4, 973

\bibitem[\protect\citeauthoryear{Farmer \& Sidorowich}{Farmer \&
  Sidorowich}{1987}]{Farmer}
Farmer J.~D.,  Sidorowich J.~J.,  1987, \mn@doi [Phys. Rev. Lett.]
  {10.1103/PhysRevLett.59.845}, 59, 845

\bibitem[\protect\citeauthoryear{{Fiocchi}, {Bazzano}, {Ubertini}  \&
  {Zdziarski}}{{Fiocchi} et~al.}{2007}]{Fiocchi}
{Fiocchi} M.,  {Bazzano} A.,  {Ubertini} P.,   {Zdziarski} A.~A.,  2007,
  \mn@doi [\apj] {10.1086/510573}, \href
  {http://adsabs.harvard.edu/abs/2007ApJ...657..448F} {657, 448}

\bibitem[\protect\citeauthoryear{{Fisher}}{{Fisher}}{1929}]{Fisher}
{Fisher} R.~A.,  1929, \mn@doi [Proceedings of the Royal Society of London A:
  Mathematical, Physical and Engineering Sciences] {10.1098/rspa.1929.0151},
  125, 54

\bibitem[\protect\citeauthoryear{{Foulkes}, {Haswell}  \& {Murray}}{{Foulkes}
  et~al.}{2006}]{Foulkes1}
{Foulkes} S.~B.,  {Haswell} C.~A.,   {Murray} J.~R.,  2006, \mn@doi [\mnras]
  {10.1111/j.1365-2966.2005.09910.x}, \href
  {http://adsabs.harvard.edu/abs/2006MNRAS.366.1399F} {366, 1399}

\bibitem[\protect\citeauthoryear{{Foulkes}, {Haswell}  \& {Murray}}{{Foulkes}
  et~al.}{2010}]{Foulkes2}
{Foulkes} S.~B.,  {Haswell} C.~A.,   {Murray} J.~R.,  2010, \mn@doi [\mnras]
  {10.1111/j.1365-2966.2009.15721.x}, \href
  {http://adsabs.harvard.edu/abs/2010MNRAS.401.1275F} {401, 1275}

\bibitem[\protect\citeauthoryear{Gao, Cao, wen Tung  \& Hu}{Gao
  et~al.}{2007}]{Gao}
Gao J.,  Cao Y.,  wen Tung W.,   Hu J.,  2007, Multiscale Analysis of Complex
  Time Series: Integration of Chaos and Random Fractal Theory, and Beyond.
Wiley-Interscience

\bibitem[\protect\citeauthoryear{{Gilmore}}{{Gilmore}}{1998}]{Gilmore}
{Gilmore} R.,  1998, \mn@doi [Reviews of Modern Physics]
  {10.1103/RevModPhys.70.1455}, \href
  {http://adsabs.harvard.edu/abs/1998RvMP...70.1455G} {70, 1455}

\bibitem[\protect\citeauthoryear{Gilmore}{Gilmore}{2007}]{GilmoreBook}
Gilmore R.~S.,  2007, The symmetry of chaos.
Oxford University Press, Oxford New York

\bibitem[\protect\citeauthoryear{Gilmore \& Lefranc}{Gilmore \&
  Lefranc}{2002}]{GilmoreTopologyBook}
Gilmore R.,  Lefranc M.,  2002, The Topology of Chaos: Alice in Stretch and
  Squeezeland.
Wiley-VCH

\bibitem[\protect\citeauthoryear{Grzedzielski, Sukova  \& Janiuk}{Grzedzielski
  et~al.}{2015}]{Grzedzielski}
Grzedzielski M.,  Sukova P.,   Janiuk A.,  2015, \mn@doi [Journal of
  Astrophysics and Astronomy] {10.1007/s12036-015-9356-7}, 36, 0

\bibitem[\protect\citeauthoryear{{Hasinger} \& {van der Klis}}{{Hasinger} \&
  {van der Klis}}{1989}]{Hasinger}
{Hasinger} G.,  {van der Klis} M.,  1989, \aap, \href
  {http://adsabs.harvard.edu/abs/1989A%26A...225...79H} {225, 79}

\bibitem[\protect\citeauthoryear{Hegger, Kantz  \& Schreiber}{Hegger
  et~al.}{1999}]{Tisean}
Hegger R.,  Kantz H.,   Schreiber T.,  1999, \mn@doi [Chaos: An
  Interdisciplinary Journal of Nonlinear Science] {10.1063/1.166424}, 9, 413

\bibitem[\protect\citeauthoryear{{Hilton}}{{Hilton}}{2011}]{Hilton}
{Hilton} E.~J.,  2011, PhD thesis, University of Washington

\bibitem[\protect\citeauthoryear{Holmes \& Rand}{Holmes \& Rand}{1980}]{Holmes}
Holmes P.,  Rand D.,  1980, \mn@doi [International Journal of Non-Linear
  Mechanics] {https://doi.org/10.1016/0020-7462(80)90031-1}, 15, 449

\bibitem[\protect\citeauthoryear{{Homan}, {Kaplan}, {van den Berg}  \&
  {Young}}{{Homan} et~al.}{2009}]{Homan}
{Homan} J.,  {Kaplan} D.~L.,  {van den Berg} M.,   {Young} A.~J.,  2009,
  \mn@doi [\apj] {10.1088/0004-637X/692/1/73}, \href
  {http://adsabs.harvard.edu/abs/2009ApJ...692...73H} {692, 73}

\bibitem[\protect\citeauthoryear{Hope}{Hope}{1968}]{Hope}
Hope A. C.~A.,  1968, Journal of the Royal Statistical Society. Series B
  (Methodological), 30, 582

\bibitem[\protect\citeauthoryear{Kanamaru}{Kanamaru}{2008}]{DuffFigure}
Kanamaru T.,  2008, \mn@doi [Scholarpedia] {10.4249/scholarpedia.6327}, 3, 6327

\bibitem[\protect\citeauthoryear{Kantz \& Schreiber}{Kantz \&
  Schreiber}{2004}]{Kantz}
Kantz H.,  Schreiber T.,  2004, Nonlinear Time Series Analysis.
Cambridge University Press

\bibitem[\protect\citeauthoryear{{Kirsch} et~al.,}{{Kirsch}
  et~al.}{2005}]{Kirsch}
{Kirsch} M.~G.,  et~al., 2005, in {Siegmund} O.~H.~W.,  ed.,  \procspie Vol.
  5898, UV, X-Ray, and Gamma-Ray Space Instrumentation for Astronomy XIV. pp
  22--33 (\mn@eprint {} {astro-ph/0508235}), \mn@doi{10.1117/12.616893}

\bibitem[\protect\citeauthoryear{Lai}{Lai}{1999}]{Lai}
Lai D.,  1999, The Astrophysical Journal, 524, 1030

\bibitem[\protect\citeauthoryear{Landau \& Lifshitz}{Landau \&
  Lifshitz}{2013}]{Landau}
Landau L.,  Lifshitz E.,  2013, Course of Theoretical Physics.
Elsevier Science

\bibitem[\protect\citeauthoryear{Letellier, Roulin  \& Rössler}{Letellier
  et~al.}{2006}]{Letellier}
Letellier C.,  Roulin E.,   Rössler O.~E.,  2006, \mn@doi [Chaos, Solitons &
  Fractals] {https://doi.org/10.1016/j.chaos.2005.05.036}, 28, 337

\bibitem[\protect\citeauthoryear{{Levine}, {Bradt}, {Cui}, {Jernigan},
  {Morgan}, {Remillard}, {Shirey}  \& {Smith}}{{Levine} et~al.}{1996}]{Levine}
{Levine} A.~M.,  {Bradt} H.,  {Cui} W.,  {Jernigan} J.~G.,  {Morgan} E.~H.,
  {Remillard} R.,  {Shirey} R.~E.,   {Smith} D.~A.,  1996, \mn@doi [\apjl]
  {10.1086/310260}, \href {http://adsabs.harvard.edu/abs/1996ApJ...469L..33L}
  {469, L33}

\bibitem[\protect\citeauthoryear{{Levine}, {Bradt}, {Chakrabarty}, {Corbet}  \&
  {Harris}}{{Levine} et~al.}{2011}]{Levine2011}
{Levine} A.~M.,  {Bradt} H.~V.,  {Chakrabarty} D.,  {Corbet} R.~H.~D.,
  {Harris} R.~J.,  2011, \mn@doi [\apjs] {10.1088/0067-0049/196/1/6}, \href
  {http://adsabs.harvard.edu/abs/2011ApJS..196....6L} {196, 6}

\bibitem[\protect\citeauthoryear{Li \& Yorke}{Li \&
  Yorke}{1975}]{PeriodThree_Li}
Li T.-Y.,  Yorke J.~A.,  1975, The American Mathematical Monthly, 82, 985

\bibitem[\protect\citeauthoryear{{Lin}, {Remillard}  \& {Homan}}{{Lin}
  et~al.}{2010}]{Lin}
{Lin} D.,  {Remillard} R.~A.,   {Homan} J.,  2010, \mn@doi [\apj]
  {10.1088/0004-637X/719/2/1350}, \href
  {http://adsabs.harvard.edu/abs/2010ApJ...719.1350L} {719, 1350}

\bibitem[\protect\citeauthoryear{Lorenz}{Lorenz}{1963}]{Lorenz}
Lorenz E.~N.,  1963, \mn@doi [Journal of the Atmospheric Sciences]
  {10.1175/1520-0469(1963)020<0130:dnf>2.0.co;2}, 20, 130

\bibitem[\protect\citeauthoryear{Marwan, Romano, Thiel  \& Kurths}{Marwan
  et~al.}{2007}]{Marwan}
Marwan N.,  Romano M.~C.,  Thiel M.,   Kurths J.,  2007, \mn@doi [Physics
  Reports] {https://doi.org/10.1016/j.physrep.2006.11.001}, 438, 237

\bibitem[\protect\citeauthoryear{{Matsuoka} et~al.,}{{Matsuoka}
  et~al.}{2009}]{Matsuoka}
{Matsuoka} M.,  et~al., 2009, \mn@doi [\pasj] {10.1093/pasj/61.5.999}, \href
  {http://adsabs.harvard.edu/abs/2009PASJ...61..999M} {61, 999}

\bibitem[\protect\citeauthoryear{{Mindlin} \& {Gilmore}}{{Mindlin} \&
  {Gilmore}}{1992}]{Mindlin2}
{Mindlin} G.~M.,  {Gilmore} R.,  1992, \mn@doi [Physica D Nonlinear Phenomena]
  {10.1016/0167-2789(92)90111-Y}, \href
  {http://adsabs.harvard.edu/abs/1992PhyD...58..229M} {58, 229}

\bibitem[\protect\citeauthoryear{{Mindlin}, {Hou}, {Solari}, {Gilmore}  \&
  {Tufillaro}}{{Mindlin} et~al.}{1990}]{Mindlin1}
{Mindlin} G.~B.,  {Hou} X.-J.,  {Solari} H.~G.,  {Gilmore} R.,   {Tufillaro}
  N.~B.,  1990, \mn@doi [Physical Review Letters]
  {10.1103/PhysRevLett.64.2350}, \href
  {http://adsabs.harvard.edu/abs/1990PhRvL..64.2350M} {64, 2350}

\bibitem[\protect\citeauthoryear{Mindlin, Solari, Natiello, Gilmore  \&
  J.~Hou}{Mindlin et~al.}{1991}]{Mindlin3}
Mindlin G.,  Solari H.,  Natiello M.,  Gilmore R.,   J.~Hou X.,  1991, 1, 147

\bibitem[\protect\citeauthoryear{{Morsink} \& {Stella}}{{Morsink} \&
  {Stella}}{1999}]{Morsink}
{Morsink} S.~M.,  {Stella} L.,  1999, \mn@doi [\apj] {10.1086/306876}, \href
  {http://adsabs.harvard.edu/abs/1999ApJ...513..827M} {513, 827}

\bibitem[\protect\citeauthoryear{{Muno}, {Remillard}  \& {Chakrabarty}}{{Muno}
  et~al.}{2002}]{Muno}
{Muno} M.~P.,  {Remillard} R.~A.,   {Chakrabarty} D.,  2002, \mn@doi [\apjl]
  {10.1086/340269}, \href {http://adsabs.harvard.edu/abs/2002ApJ...568L..35M}
  {568, L35}

\bibitem[\protect\citeauthoryear{{Ogilvie} \& {Dubus}}{{Ogilvie} \&
  {Dubus}}{2001}]{Ogilvie2}
{Ogilvie} G.~I.,  {Dubus} G.,  2001, \mn@doi [\mnras]
  {10.1046/j.1365-8711.2001.04011.x}, \href
  {http://adsabs.harvard.edu/abs/2001MNRAS.320..485O} {320, 485}

\bibitem[\protect\citeauthoryear{{Olive}, {Barret}  \&
  {Gierli{\'n}ski}}{{Olive} et~al.}{2003}]{Olive}
{Olive} J.-F.,  {Barret} D.,   {Gierli{\'n}ski} M.,  2003, \mn@doi [\apj]
  {10.1086/344835}, \href {http://adsabs.harvard.edu/abs/2003ApJ...583..416O}
  {583, 416}

\bibitem[\protect\citeauthoryear{{Osborne} \& {Provenzale}}{{Osborne} \&
  {Provenzale}}{1989}]{Osborne}
{Osborne} A.~R.,  {Provenzale} A.,  1989, \mn@doi [Physica D Nonlinear
  Phenomena] {10.1016/0167-2789(89)90075-4}, \href
  {http://adsabs.harvard.edu/abs/1989PhyD...35..357O} {35, 357}

\bibitem[\protect\citeauthoryear{Ott}{Ott}{2002}]{Ott}
Ott E.,  2002, Chaos in Dynamical Systems, 2 edn.
Cambridge University Press, \mn@doi{10.1017/CBO9780511803260}

\bibitem[\protect\citeauthoryear{{Piraino}, {Santangelo}, {di Salvo}, {Kaaret},
  {Horns}, {Iaria}  \& {Burderi}}{{Piraino} et~al.}{2007}]{Piraino}
{Piraino} S.,  {Santangelo} A.,  {di Salvo} T.,  {Kaaret} P.,  {Horns} D.,
  {Iaria} R.,   {Burderi} L.,  2007, \mn@doi [\aap]
  {10.1051/0004-6361:20077841}, \href
  {http://adsabs.harvard.edu/abs/2007A%26A...471L..17P} {471, L17}

\bibitem[\protect\citeauthoryear{{Piraino}, {Santangelo}, {M{\"u}ck}, {Kaaret},
  {Di Salvo}, {D'A{\`i}}, {Iaria}  \& {Egron}}{{Piraino}
  et~al.}{2016}]{Piraino2016}
{Piraino} S.,  {Santangelo} A.,  {M{\"u}ck} B.,  {Kaaret} P.,  {Di Salvo} T.,
  {D'A{\`i}} A.,  {Iaria} R.,   {Egron} E.,  2016, \mn@doi [\aap]
  {10.1051/0004-6361/201424150}, \href
  {http://adsabs.harvard.edu/abs/2016A%26A...591A..41P} {591, A41}

\bibitem[\protect\citeauthoryear{Poincar\'e}{Poincar\'e}{2010}]{PoincareBook}
Poincar\'e H.,  2010, Papers on Topology: Analysis Situs and Its Five
  Supplements (History of Mathematics).
American Mathematical Society

\bibitem[\protect\citeauthoryear{{Postnov}, {Shakura}, {Staubert},
  {Kochetkova}, {Klochkov}  \& {Wilms}}{{Postnov} et~al.}{2013}]{Postnov}
{Postnov} K.,  {Shakura} N.,  {Staubert} R.,  {Kochetkova} A.,  {Klochkov} D.,
   {Wilms} J.,  2013, \mn@doi [\mnras] {10.1093/mnras/stt1363}, \href
  {http://adsabs.harvard.edu/abs/2013MNRAS.435.1147P} {435, 1147}

\bibitem[\protect\citeauthoryear{{Pringle}}{{Pringle}}{1996}]{Pringle96}
{Pringle} J.~E.,  1996, \mn@doi [\mnras] {10.1093/mnras/281.1.357}, \href
  {http://adsabs.harvard.edu/abs/1996MNRAS.281..357P} {281, 357}

\bibitem[\protect\citeauthoryear{R\"{o}ssler}{R\"{o}ssler}{1976}]{Rossler1976}
R\"{o}ssler O.,  1976, \mn@doi [Physics Letters A]
  {10.1016/0375-9601(76)90101-8}, 57, 397

\bibitem[\protect\citeauthoryear{{R{\"o}ssler}}{{R{\"o}ssler}}{1977}]{Rossler1977}
{R{\"o}ssler} O.~E.,  1977, \mn@doi [Physics Letters A]
  {10.1016/0375-9601(77)90029-9}, \href
  {http://adsabs.harvard.edu/abs/1977PhLA...60..392R} {60, 392}

\bibitem[\protect\citeauthoryear{Sauer}{Sauer}{1994}]{Sauer94}
Sauer T.,  1994, in Weigend A.~S.,  Gershenfeld N.~A.,  eds, , Time Series
  Prediction: Forecasting the Future and Understanding the Past.
Addison-Wesley

\bibitem[\protect\citeauthoryear{Schreiber \& Schmitz}{Schreiber \&
  Schmitz}{1996}]{Schreiber1996}
Schreiber T.,  Schmitz A.,  1996, \mn@doi [Phys. Rev. Lett.]
  {10.1103/PhysRevLett.77.635}, 77, 635

\bibitem[\protect\citeauthoryear{Schreiber \& Schmitz}{Schreiber \&
  Schmitz}{2000}]{Schreiber2000}
Schreiber T.,  Schmitz A.,  2000, \mn@doi [Physica D: Nonlinear Phenomena]
  {10.1016/s0167-2789(00)00043-9}, 142, 346

\bibitem[\protect\citeauthoryear{{Shapiro} \& {Teukolsky}}{{Shapiro} \&
  {Teukolsky}}{1983}]{Shapiro}
{Shapiro} S.~L.,  {Teukolsky} S.~A.,  1983, {Black holes, white dwarfs, and
  neutron stars: The physics of compact objects}.
Wiley, New York

\bibitem[\protect\citeauthoryear{Shimshoni}{Shimshoni}{1971}]{Shimshoni}
Shimshoni M.,  1971, \mn@doi [Geophysical Journal of the Royal Astronomical
  Society] {10.1111/j.1365-246X.1971.tb01829.x}, 23, 373

\bibitem[\protect\citeauthoryear{{Smale} \& {Boyd}}{{Smale} \&
  {Boyd}}{2012}]{Smale}
{Smale} A.~P.,  {Boyd} P.~T.,  2012, \mn@doi [\apj]
  {10.1088/0004-637X/756/2/146}, \href
  {http://adsabs.harvard.edu/abs/2012ApJ...756..146S} {756, 146}

\bibitem[\protect\citeauthoryear{{Solari} \& {Gilmore}}{{Solari} \&
  {Gilmore}}{1988a}]{Solari1}
{Solari} H.~G.,  {Gilmore} R.,  1988a, \mn@doi [\pra]
  {10.1103/PhysRevA.37.3096}, \href
  {http://adsabs.harvard.edu/abs/1988PhRvA..37.3096S} {37, 3096}

\bibitem[\protect\citeauthoryear{{Solari} \& {Gilmore}}{{Solari} \&
  {Gilmore}}{1988b}]{Solari2}
{Solari} H.~G.,  {Gilmore} R.,  1988b, \mn@doi [\pra]
  {10.1103/PhysRevA.38.1566}, \href
  {http://adsabs.harvard.edu/abs/1988PhRvA..38.1566S} {38, 1566}

\bibitem[\protect\citeauthoryear{{Still} \& {Boyd}}{{Still} \&
  {Boyd}}{2004}]{Still}
{Still} M.,  {Boyd} P.,  2004, \mn@doi [\apjl] {10.1086/421349}, \href
  {http://adsabs.harvard.edu/abs/2004ApJ...606L.135S} {606, L135}

\bibitem[\protect\citeauthoryear{{Sukov{\'a}}, {Grzedzielski}  \&
  {Janiuk}}{{Sukov{\'a}} et~al.}{2016}]{Sukova}
{Sukov{\'a}} P.,  {Grzedzielski} M.,   {Janiuk} A.,  2016, \mn@doi [\aap]
  {10.1051/0004-6361/201526692}, \href
  {http://adsabs.harvard.edu/abs/2016A%26A...586A.143S} {586, A143}

\bibitem[\protect\citeauthoryear{{Takens}}{{Takens}}{1981}]{Takens}
{Takens} F.,  1981, \mn@doi [Lecture Notes in Mathematics, Berlin Springer
  Verlag] {10.1007/BFb0091924}, \href
  {http://adsabs.harvard.edu/abs/1981LNM...898..366T} {898, 366}

\bibitem[\protect\citeauthoryear{Theiler \& Prichard}{Theiler \&
  Prichard}{1996}]{Theiler1996}
Theiler J.,  Prichard D.,  1996, \mn@doi [Physica D: Nonlinear Phenomena]
  {https://doi.org/10.1016/0167-2789(96)00050-4}, 94, 221

\bibitem[\protect\citeauthoryear{Theiler, Eubank, Longtin, Galdrikian  \&
  Farmer}{Theiler et~al.}{1992}]{Theiler1992}
Theiler J.,  Eubank S.,  Longtin A.,  Galdrikian B.,   Farmer J.~D.,  1992,
  \mn@doi [Physica D: Nonlinear Phenomena] {10.1016/0167-2789(92)90102-s}, 58,
  77

\bibitem[\protect\citeauthoryear{{Torpin}, {Boyd}, {Smale}  \&
  {Valencic}}{{Torpin} et~al.}{2017}]{Torpin}
{Torpin} T.~J.,  {Boyd} P.~T.,  {Smale} A.~P.,   {Valencic} L.~A.,  2017,
  \mn@doi [\apj] {10.3847/1538-4357/aa8f96}, \href
  {http://adsabs.harvard.edu/abs/2017ApJ...849...32T} {849, 32}

\bibitem[\protect\citeauthoryear{{Tufillaro}, {Solari}  \&
  {Gilmore}}{{Tufillaro} et~al.}{1990}]{Tufillaro}
{Tufillaro} N.~B.,  {Solari} H.~G.,   {Gilmore} R.,  1990, \mn@doi [\pra]
  {10.1103/PhysRevA.41.5717}, \href
  {http://adsabs.harvard.edu/abs/1990PhRvA..41.5717T} {41, 5717}

\bibitem[\protect\citeauthoryear{{Wang}, {Chang}, {Zhang}, {Wang}, {Chen}, {Qu}
   \& {Song}}{{Wang} et~al.}{2012}]{Wang}
{Wang} J.,  {Chang} H.~K.,  {Zhang} C.~M.,  {Wang} D.~H.,  {Chen} L.,  {Qu}
  J.~L.,   {Song} L.~M.,  2012, \mn@doi [\apss] {10.1007/s10509-012-1173-8},
  \href {http://adsabs.harvard.edu/abs/2012Ap%26SS.342..357W} {342, 357}

\bibitem[\protect\citeauthoryear{{Whitmire} \& {Matese}}{{Whitmire} \&
  {Matese}}{1980}]{Whitmire}
{Whitmire} D.~P.,  {Matese} J.~J.,  1980, \mn@doi [\mnras]
  {10.1093/mnras/193.4.707}, \href
  {http://adsabs.harvard.edu/abs/1980MNRAS.193..707W} {193, 707}

\bibitem[\protect\citeauthoryear{{Wojdowski}, {Clark}, {Levine}, {Woo}  \&
  {Zhang}}{{Wojdowski} et~al.}{1998}]{Wojdowski}
{Wojdowski} P.,  {Clark} G.~W.,  {Levine} A.~M.,  {Woo} J.~W.,   {Zhang} S.~N.,
   1998, \mn@doi [\apj] {10.1086/305893}, \href
  {http://adsabs.harvard.edu/abs/1998ApJ...502..253W} {502, 253}

\bibitem[\protect\citeauthoryear{Zunino, Soriano  \& Rosso}{Zunino
  et~al.}{2012}]{Zunino}
Zunino L.,  Soriano M.~C.,   Rosso O.~A.,  2012, \mn@doi [Phys. Rev. E]
  {10.1103/PhysRevE.86.046210}, 86, 046210

\bibitem[\protect\citeauthoryear{{van der Klis}}{{van der
  Klis}}{2006}]{vanDerKlis}
{van der Klis} M.,  2006, {Rapid X-ray Variability}.
Cambridge University Press, Cambridge, UK, pp 39--112

\makeatother
\end{thebibliography}

\appendix
\section{The Method of Surrogate Data}
The application of non-linear time series methods should be motivated by evidence of non-linearity in the time series in question. The method of surrogate data testing introduced by \cite{Theiler1992} attempts to find the 'least interesting' explanation that cannot be ruled out based on the data, namely that some type of linear stochastic process is responsible for the observables of interest. In summary, we do this by first formulating a suitable null hypothesis for the observed structures in the time series. We then generate 'surrogate' data that preserves the temporal correlations and probability distribution of the original data using a Fourier-based Monte Carlo resampling technique. We can look for additional structure that is present in the data but not in the surrogates via statistical tests, thereby indicating non-linearity and possibly determinism, depending on the nature of the test statistic used.

The method of surrogate data is different than the statistical approach of choosing model equations fit to the data, producing simulated time series from the chosen model, and then comparing to the original dataset, called the 'typical realisations' approach \citep{Theiler1996}. Instead, we follow a 'constrained realisations' approach \citep{Theiler1996} where we produce simulated time series, or surrogates, which are generated by the original dataset itself, thereby imposing the features of interest onto the surrogates whilst eliminating dynamical information, rather than fitting model equations to the data.

\begin{figure}[h]
\centering
\subfloat[]{
	\includegraphics[width=0.5\textwidth]{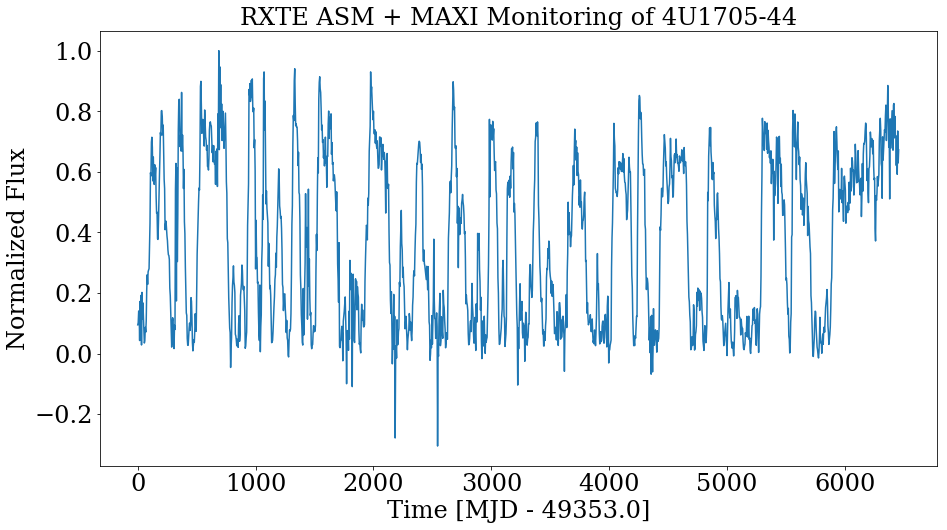}}
\qquad
\subfloat[]{
	\includegraphics[width=0.5\textwidth]{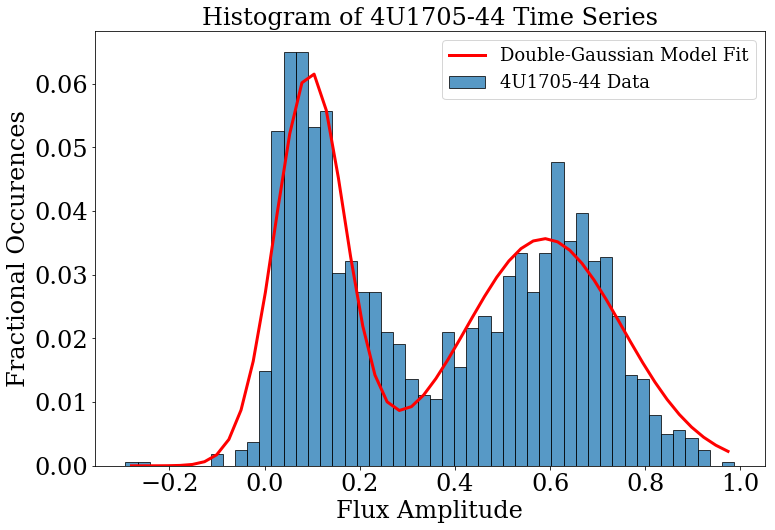}}
\qquad
\subfloat[]{
 	\includegraphics[width=.5\textwidth]{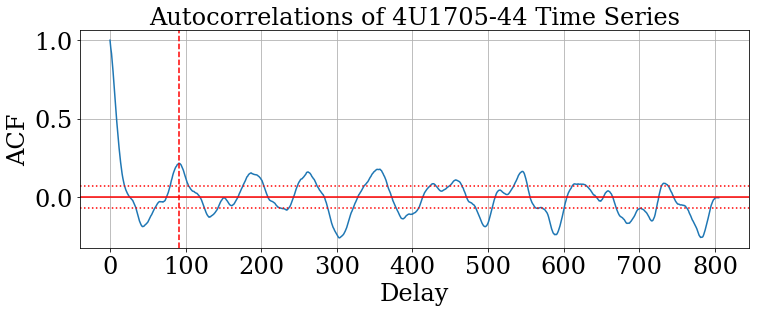}}
\caption{(a) The original time series of 4U1705--44, unfiltered and normalised to unity. (b) The distribution of 4U1705--44 normalised flux. We have fit a double gaussian model to the histogram in order to accentuate the double-welled features in the time series and for comparison to the surrogates. The fit parameters corresponding to the mean, standard deviation, and amplitude of the first and second Gaussian peaks are $(0.095\pm0.005, 0.073\pm0.005,  0.061\pm0.004, 0.59\pm0.01, 0.16\pm0.01, 0.035\pm0.003)$, where the errors are the standard deviations to the fit parameters. (c) The autocorrelation function of the time series up to a lag half the length of the original light curve. The first (also the largest) peak occurs at a delay of $\tau = 91$, corresponding to 364 days (vertical dashed line). The dotted lines correspond to 95 per cent confidence level of significance.}
\label{fig:orig_data}
\end{figure}

We seek to reproduce the bimodal behaviour (the low- and high-flux states) and the strong recurrence phenomena (the non-periodic recurrence of high and low states) present in the 4U1705--44 light curve and phase-space projection. Our null hypothesis is that a variation of a linear stochastic process produces these observed features (which variation is specified by the chosen test statistic). Our original and unfiltered 4U1705--44 light curve, normalised to unity, is plotted in Fig.~\ref{fig:orig_data}(a). The distribution, or histogram, of the 4U1705--44 time series, Fig.~\ref{fig:orig_data}(b), reflects the amount of time, fractionally, that the light curve is spent in a specific flux bin and clearly distinguishes the two peaks corresponding to the high- and low-flux states. The \textit{correlogram}, or autocorrelation function, of 4U1705--44 is computed up to a time delay of half the length of the time series in Fig.~\ref{fig:orig_data}(c). The autocorrelation function is defined as:
\begin{equation}
	C(\tau) = \frac{1}{N-\tau} \frac{\sum_i (x_i - y) (x_{i+\tau} - y)}{\sigma^2(x)}
\end{equation}
where $\tau$ is the time delay or lag, $N$ is the total number of data points, $y$ is the average, $\sigma$ is the standard deviation, and $x$ is the time series \citep{Tisean}. That is, we have plotted the correlation of the time series to a delayed copy of itself as a function of the delay. The first peak, which is also the most positive correlation, corresponds to a time delay of 364 days. This timescale is on the order of the period-3 extracted from the 4U1705--44 light curve in the Close Returns analysis in this paper. It is worth noting that strong period-3 occurrences are indicators of chaos \citep{PeriodThree_Li}. We also observe a very slow decay overall in the autocorrelation, with multiple significant correlated and anti-correlated peaks (above or below the dotted red lines of 95 per cent confidence levels in Fig.~\ref{fig:orig_data}(c)), signifying the presence of both long-term memory and periodic components.

\subsection{The Surrogate Data}
Using the time series analysis code, \textit{TISEAN} (\citealt{Tisean}, \citealt{Schreiber2000}), we create a set of surrogate data whereby an iterative algorithm (the routine called \textit{surrogates}) ensures the surrogates obtain the same autocorrelations as the data and the same probability distribution \citep{Schreiber1996}, the features of interest as in Fig.~\ref{fig:orig_data}. To produce our confidence level, we use a rank-order test, as introduced by \cite{Theiler1992} and summarised by \cite{Schreiber2000}, avoiding the likely misplaced assumption that our signal has a single Gaussian distribution. We can obtain a level of significance of $(1-\alpha)\times 100\%$, where $\alpha$ is the residual probability of a false rejection, by generating $M = 2K/\alpha - 1$ surrogate datasets where $K$ is an integer resulting in a probability $\alpha$ that the data gives either one of the $K$ smallest or largest values (see \cite{Hope} for an extensive proof).

\begin{figure}
\centering
\subfloat[]{
	\includegraphics[width=0.5\textwidth]{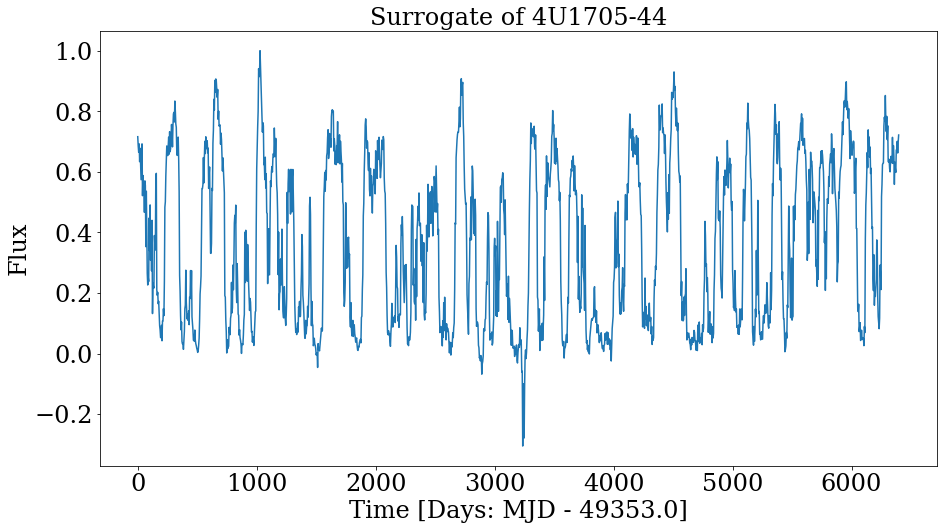}}
\qquad
\subfloat[]{
	\includegraphics[width=0.5\textwidth]{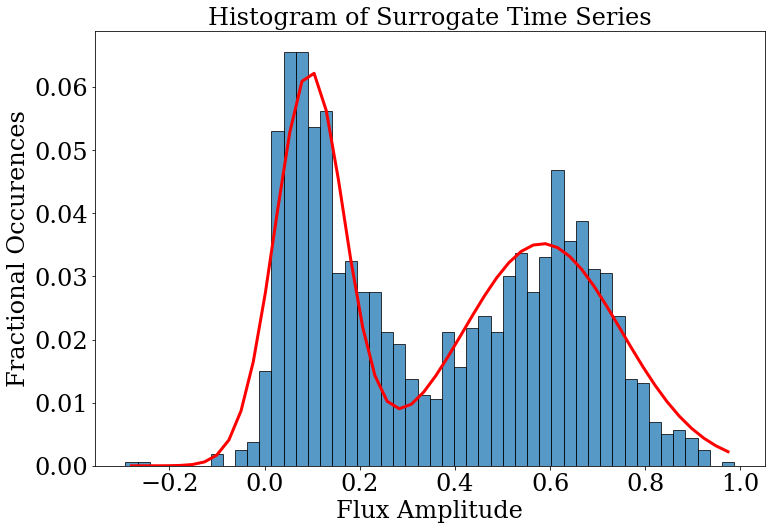}}
\qquad
\subfloat[]{
 	\includegraphics[width=.5\textwidth]{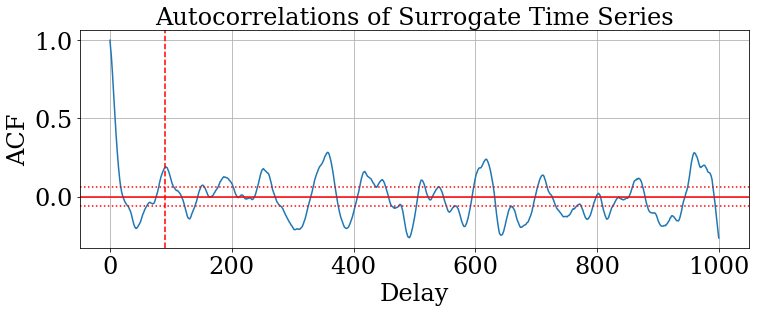}}
\caption{Same as Fig.~\ref{fig:orig_data}. The distribution and autocorrelations are conserved from the original dataset. Namely, in (b), the mean, standard deviation, and amplitude of the 1st and 2nd peaks are, respectively, $(0.095\pm0.005,  0.072\pm0.005,  0.062\pm0.004,  0.58\pm0.01, 0.17\pm0.01,  0.035\pm0.002)$, which are within the standard deviation of the Fig.~\ref{fig:orig_data}(b) parameters. In (c), the first peak (dashed line) is significant, occurring above the 95 per cent confidence level (dotted line), and occurs at the same delay coordinate of $\tau = 91$, or 364 days.}
\label{fig:surr_data}
\end{figure}

We desire a significance level of at least 95 per cent and choose $K=2$ (generally chosen to be a small integer in order to reduce computation time), requiring 79 surrogate time series. Whichever test statistic we use will require the original time series statistic to fall in the top or bottom $K=2$ values for a 95 per cent level of significance. An example surrogate time series is plotted in Fig.~\ref{fig:surr_data} alongside its distribution and \textit{correlogram} where we note that both are conserved between the original time series and the generated surrogate.

The test design for generating Fourier based surrogates and other methods and their associated confidence levels are described by \cite{Schreiber2000} and the corresponding C and Fortran based computational methods for surrogate generation and testing are available in the \textit{TISEAN} package \citep{Tisean}.

\subsection{Test Statistics}
\begin{figure}
	\centering
	\includegraphics[width=0.5\textwidth]{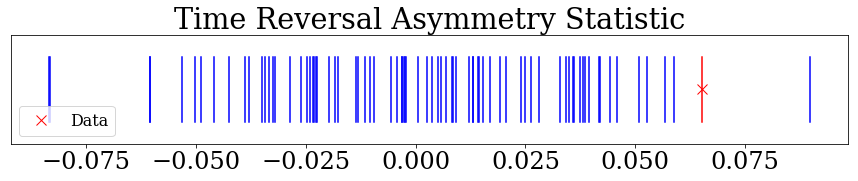}
	\caption{Time reversibility asymmetry statistic as defined by Eq.~\ref{eq:timerev} for the data, marked by the red line and 'x', and the 79 surrogates. As the time reversibility of the 4U1705--44 data is one of the two greatest or least, we can rule out the null hypothesis of any transformation of a linear Gaussian stochastic process with a 95 per cent level of significance.}
	\label{fig:timerev}
\end{figure}

There are several test statistics that can detect non-linearity and tracers of chaos to provide motivation for pursuing the non-linear topological approach in this paper. A test often used for detecting non-linearity in a time series is the time reversal asymmetry statistic. That is, a system that is non-linear (deterministic or stochastic) will lead to time irreversibility, or poor forecasting of the time-reversed data. \cite{Diks} showed that if reversibility can be rejected, all static transformations of linear stochastic processes can be excluded as a model for the original time series. Such a distinction is important because the power spectrum does not contain information about the direction of time and is therefore insufficient for distinguishing linear stochastic and other related processes. The \textit{TISEAN} package includes the function \textit{timerev} for the purpose of testing time reversibility, which \cite{Schreiber2000} defines quantitatively as:
\begin{equation}\label{eq:timerev}
	\phi^{rev}(\tau) = \frac{1}{N-\tau} \sum_{n=\tau + 1}^{N} (s_n - s_{n-\tau})^3,
\end{equation}
where $\tau$ represents the time delay and $N$ the length of the time series, $s_N$.
We compute the time reversibility asymmetry statistic for the data and the 79 surrogates with the results plotted in Fig.~\ref{fig:timerev}, in which the time asymmetry of the original data is found to be significantly different from that of the surrogates. That is, the null hypothesis of any static transformation of a linear stochastic process generating the time series of 4U1705--44 can be rejected at a 95 per cent significance level, since the resulting statistic for the data is found to be either the first or the last two values against the surrogates. It should be noted, however, that time reversal asymmetry detects non-linearity, but not the nature of the non-linearity, i.e. deterministic or stochastic.

Another method to test the null hypothesis of a stationary, possibly rescaled, linear Gaussian random process generating the observed time series is via a simple non-linear prediction algorithm \citep{Kantz}, which was originally proposed as a forecasting method by \cite{Lorenz} referred to as the "Lorenz' method of analogues." This method proposes that, if we have observed a system for some time, there will be states in the past which are arbitrarily close to the current state, in which a 'state' is considered to be some segment of the time series, embedded in phase space. We can therefore use a localised segment of the embedded time series to predict, or forecast, future behaviour.

Given a time series $y$ of length $N$ and its corresponding $m$-dimensional embedding in phase space, one searches for all embedding vectors, $s_n$, in the past which are closer than some $\epsilon$ distance to the current state, $s_N$ (i.e. for the last segment in the time series, we search for every other previous segment in the time series which follows a similar time evolution in phase space). The $m\times\epsilon$ region centred about $s_n$ in phase-space forms a neighbourhood, $U_{\epsilon}(s_N)$. One then takes the average of all regions, $s_n$, found to be within the neighbourhood $U_{\epsilon}(s_N)$ and uses it as the prediction for the future evolution of the time series, a distance $\Delta n$ away from the end of the time series. In the \textit{TISEAN} package, we follow the \cite{Kantz} definition of the resulting predicted state:
\begin{equation}
	\hat{s}_{N+\Delta n} = \frac{1}{\mid U_{\epsilon}(s_N) \mid} \sum_{s_n \in U_{\epsilon}(s_N)} s_{n+\Delta n}
\end{equation}
The described algorithm above is considered a zeroth-order prediction of the dynamics whereby we are using a locally constant predictor to forecast the future of the time series. Given that this method exploits deterministic structures in the signal, the expectation is that, in applying a prediction algorithm in surrogate testing, the resulting $rms$ error in the prediction algorithm \citep{Farmer} would be smallest for the original time series than in its surrogates if the null hypothesis is to be rejected \citep{Tisean}. For the case of the 4U1705--44 time series and its 79 surrogates, we require the simple non-linear prediction test error to be one of the two smallest errors in order to reject the null hypothesis with a 95 per cent level of significance. Per Fig.~\ref{fig:simple_non-linear}, this is the case.

\begin{figure}
	\centering
	\includegraphics[width=0.5\textwidth]{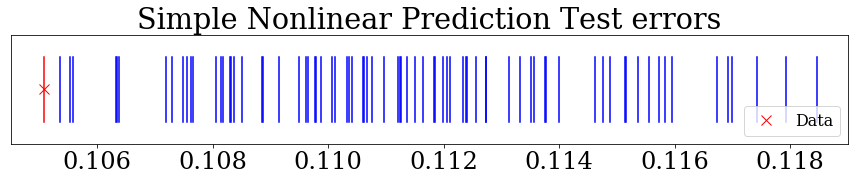}
	\caption{The $rms$ error of the simple non-linear prediction test (\citealt{Farmer}, \citealt{Tisean}) of the 4U1705--44 data against its 79 surrogates. The original data maintains one of the smallest two errors and is therefore significant at the 95 per cent level to rule out the null hypothesis of a linear stochastic process.}
	\label{fig:simple_nonlinear}
\end{figure}

\begin{figure}
	\centering
	\includegraphics[width=0.45\textwidth]{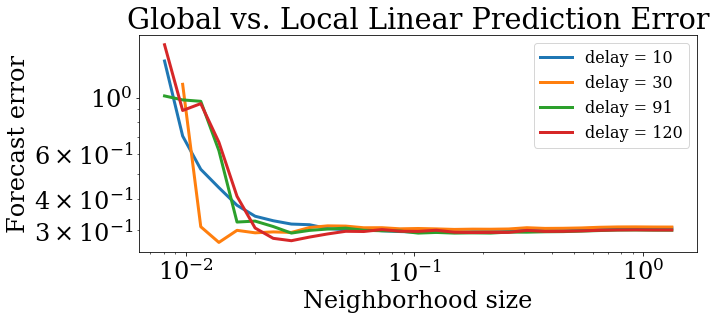}
	\caption{Local versus global prediction errors as a function of neighbourhood size, against logarithmic axes, for four different delay embeddings, corresponding to 40, 120, 364, and 460 days, respectively. The neighbourhood size is determined by the square distance about the segment of the time series used for the prediction algorithm in the embedded phase space. Given that the time series has been normalised, this means that for a neighbourhood size of $5\times10^{-1}$, the distance $\epsilon$ between predictor and predicted states is more than 70 per cent the maximum distance between any two given points in the time series.}
	\label{fig:local_v_global}
\end{figure}

The third non-linear statistic that is used for detecting non-linearity present in the 4U1705--44 time series but which can also indicate determinism is the local versus global linear prediction test as described by \cite{Casdagli1992}. By defining a neighbourhood using a local linear ansatz about the data in phase space, one can then compute the average forecast error of a locally linear model as a function of the neighbourhood size. This method is algorithmically similar to the previous simple non-linear prediction test, but with using a local linear first-order model rather than a zeroth order, constant value predictor. In this case, if the agreement with the model is best at large neighbourhood sizes (that is, the error in the forecast is smallest), then the data is best described by a linear stochastic process. If instead the agreement with the model is best at the smallest neighbourhood sizes, as exemplified by the test in Fig.~\ref{fig:local_v_global}, then a non-linear, possibly deterministic, description is preferred.

\begin{figure*}
	\centering
	\includegraphics[width=0.9\textwidth]{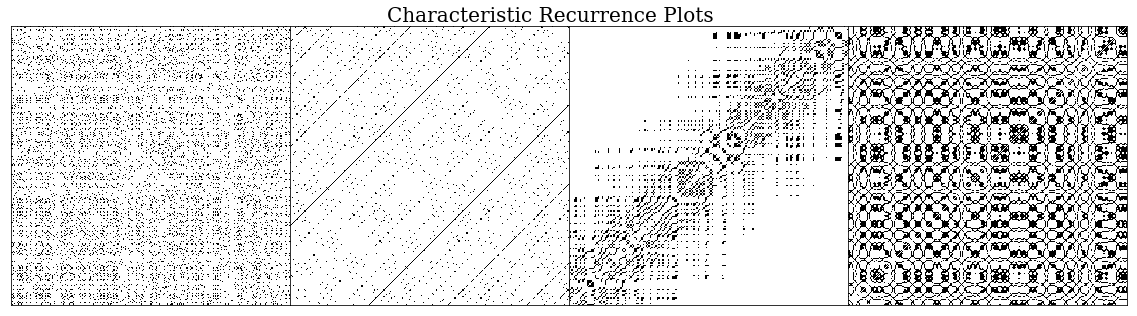}
	\caption{Four examples of recurrence plots. From left to right: uniform gaussian noise, periodic signal ($A\sin(\omega_1 t) + B\sin(\omega_2 t)$), drifting logistic map ($x_{i+1} = 4x_i(1-x_i)+0.01i$), chaotic Duffing oscillator (Eq.~\ref{eq:duff}). See similar examples with further RP structure analysis in \protect\cite{Marwan}.}
	\label{fig:recurrence_sample}
\end{figure*}

\begin{figure}
	\centering
	\includegraphics[width=0.45\textwidth]{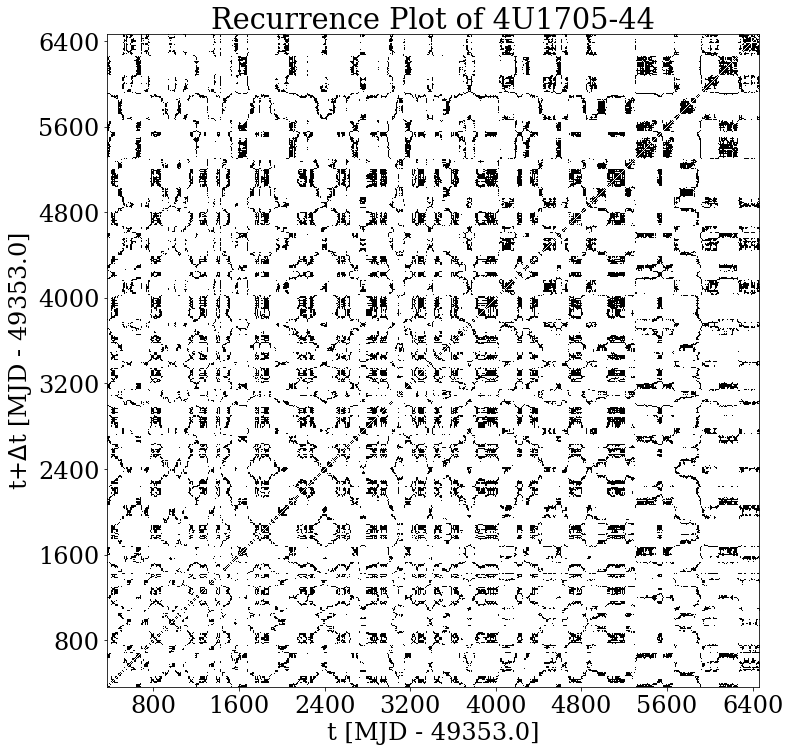}
	\caption{The recurrence plots (RP) of 4U1705--44 for a time-delay embedding of phase space using a time delay of 91 (or 364 days). A black dot is plotted when the position at $t$ is within $\epsilon = 10\% \times D$ of $t+\Delta t$, when embedded in phase space, where $D$ is the maximum phase space diameter.}
	\label{fig:recurrence_1705}
\end{figure}

\subsection{Recurrence Analysis}
The recurrence analysis tool introduced by \cite{Eckmann} and further developed by \cite{Marwan} has been used in time series analysis of various experimental data sets for the purpose of distinguishing recurrent phenomena not easily extrapolated by other correlation methods. Recurrence analysis was recently used to characterise deterministic versus stochastic structure in the light curves of black hole X-ray binaries (\citealt{Grzedzielski}, \citealt{Sukova}). The production of a recurrence plot (RP) is similar to the Close Returns method used in Sec.~\ref{sec:Period}. Every position in the time series is compared to every successive position in the time series and, if the flux-like value at those two positions are within some distance $\epsilon$ of each other in phase space, then a black dot is plotted on the RP.

Two of the parameters that must be chosen when constructing an RP are the time delay coordinate and the distance, $\epsilon$, used as a threshold for comparing two points in the time series phase space. From the \textit{correlogram} of 4U1705--44 in Fig.~\ref{fig:orig_data}, we choose a delay coordinate of 91 (or 364 days). Secondly, we follow the 'rule of thumb' outlined by \cite{Marwan} whereby the threshold $\epsilon$ should not exceed 10 per cent of the mean or the maximum phase space diameter. The resulting RP for the 4U1705--44 light curve is plotted in Fig.~\ref{fig:recurrence_1705}.

To briefly explore some of the prominent characteristics of RPs, we provide four examples of RPs from time series generated by distinctly different mechanisms. Fig.~\ref{fig:recurrence_sample} displays, in order, the RPs of uniform gaussian white noise, a periodic signal constructed from the addition of two sinusoidal waves, the logistic map corrupted by a linearly increasing term ($x_{i+1} = 4x_i(1-x_i)+0.01i$), and the chaotic Duffing oscillator (with the same parameters as the simulated time series used throughout this paper). Summarising the extensive analysis from \cite{Marwan}, $Homogeneous$ RPs, like those from gaussian white noise, are typical from stationary random processes, where no characteristic structure is distinguished perhaps because the timescales of interest are outside the bounds of the RP. Periodic or quasi-periodic systems display shorter diagonal lines parallel to the main diagonal throughout the entire RP. As noted by \cite{Marwan}, even for systems whose oscillations are not easily recognisable (like in the case for unstable periodic orbits), the presence of short diagonal lines parallel or perpendicular to the main diagonal can be useful. Non-stationary systems give rise to the fading out at the corners of the RPs, as in the case of the modified logistic map (which includes a linearly increasing term). Typical behaviour of laminar states, or intermittency, are the appearance of vertical (horizontal) lines in RPs, reflecting a period in the time series in which the state evolves slowly or does not change. Chaotic systems tend to show a combination of all of these features, except for possibly non-stationarity, as exemplified by the RP of the Duffing oscillator.

The RP of 4U1705--44 shows multiple features of interest, namely, regions of diagonal lines parallel to the main diagonal indicative of fundamental periods, short horizontal and vertical lines indicative of unstable periodic orbits and intermittency, and persistence of intensity throughout the entire RP, reflecting stationarity within the length of the time series. The qualitative similarity to the Duffing oscillator is also remarkable. At a minimum, we can state from visual inspection that there likely exists simultaneous components of noise, periodicity, and intermittent, likely chaotic components in the RP of 4U1705--44.

A more thorough and extensive analysis of the presence of non-linearity and possibly deterministic chaos can be continued on the time series 4U1705--44 using the surrogate data and recurrence analysis methods. For our purposes, we seek only a motivation to pursue non-linear analysis on the time series of 4U1705--44 using a topological comparison to the template of the Duffing oscillator, which can potentially add further justification of the presence of non-linearity and chaos if a positive correlation is found. The surrogate data testing and recurrence plot visual analysis provides ample motivation to pursue a topological analysis of the 4U1705--44 light curve as compared to the Duffing oscillator.

\label{lastpage}
\end{document}